\documentclass[twocolumn,showpacs,preprintnumbers,
amsmath,amssymb,aps,prd,nofootinbib,superscriptaddress,
eqsecnum]{revtex4}
\usepackage{graphicx,color}

\makeatletter
\renewcommand{\p@subsection}{}
\makeatother

\def\lapp{\;\raisebox{-.5ex}{$\stackrel{<}{\scriptstyle\sim}$}\;}

\renewcommand{\thesection}{\arabic{section}}

\renewcommand{\theequation}{\arabic{section}.\arabic{equation}}

\newcommand{\Slash}[1]{\ooalign{\hfil/\hfil\crcr$#1$}}

\begin{document}

\title{
Susceptibilities and the Phase Structure of a Chiral Model with  Polyakov Loops }

\author{C. Sasaki}
\affiliation{%
Gesellschaft f\"ur Schwerionenforschung, GSI,  D-64291 Darmstadt,
Germany}
\author{B. Friman}
\affiliation{%
Gesellschaft f\"ur Schwerionenforschung, GSI,  D-64291 Darmstadt,
Germany}
\author{K. Redlich}
\affiliation{%
Institute of Theoretical Physics, University of Wroclaw, PL--50204
Wroc\l aw, Poland}

\date{\today}

\begin{abstract}
In an extension of the Nambu-Jona-Lasinio model where the quarks
interact with the temporal gluon field, represented by the Polyakov
loop, we explore the relation between the deconfinement and chiral
phase transitions. The effect of Polyakov loop dynamics on
thermodynamic quantities, on the phase structure at finite
temperature and baryon density and on various susceptibilities is
presented. Particular emphasis is put on the behavior and
properties of the fluctuations of the (approximate) order
parameters and their dependence on temperature and net--quark
number density. We also discuss how the phase structure of the
model is influenced by the coupling of the quarks to the Polyakov
loop.

\end{abstract}


\setcounter{footnote}{0}

\maketitle


\section{Introduction}
\label{sec:int}

Quantum Chromodynamics (QCD) exhibits dynamical chiral symmetry
breaking  and confinement. Both features are related with global
symmetries of the QCD Lagrangian. However, the relation of
spontaneous chiral symmetry breaking and confinement still remains
an open issue. While the chiral symmetry is exact in the limit of
massless quarks the $Z(3)$ center symmetry, which governs
confinement, is exact in the opposite limit, i.e., for infinitely
heavy quarks. In order to obtain a unified picture of confinement
and chiral symmetry breaking several effective chiral models have
been studied
\cite{Gocksch,IK,Meisinger,DLS,Mocsy,Fukushima:strong,Fukushima,PNJL,Megias}.
Recently an extension of the  Nambu--Jona-Lasinio (NJL)
model~\cite{Nambu,review} was proposed and
developed~\cite{Meisinger,Fukushima,PNJL,sus:pnjl,pnjl:meson,pnjl:iso,RRW:phase}
to address this question.

The NJL model  describes  interactions of  constituent quark
fields. It exhibits a global SU(3) symmetry  that is a replacement
of a local gauge SU(3) color transformation of the QCD Lagrangian.
Thus, color confinement is lost in the NJL dynamics. Recently,
color degrees of freedom were introduced in the NJL Lagrangian
through an effective gluon potential expressed in terms of Polyakov
loops~\cite{Meisinger,Fukushima,PNJL}. Such a potential was
constructed to preserve the $Z(3)$ symmetry of the gluon part of
the QCD Lagrangian and it has its origin in the recently developed
effective models with Polyakov loops as dynamical
fields~\cite{dumitru}. An extended NJL Lagrangian (PNJL model)
contains unified properties of QCD related with $Z(3)$ and the
chiral symmetries. The interactions of quarks with the effective
gluon fields in the PNJL model is included through covariant
derivatives. Furthermore, due to the symmetries  of the Lagrangian,
the PNJL model belongs to the same universality class as that
expected for QCD. Thus, such a model can be considered as a testing
ground for studying the  phase structure and critical phenomena
related with the deconfinement and chiral phase transitions. This
is particularly interesting  since there are still limitations in
the applicability of lattice gauge theory (LGT) to QCD
thermodynamics at large net quark densities.

Recently, it was shown \cite{PNJL,sus:pnjl,pnjl:iso,RRW:phase} that
the PNJL model, formulated at finite temperature and finite quark
chemical potential, reproduces some of the thermodynamical
observables computed within LGT. The properties of the equation of
state \cite{PNJL}, the in-medium  modification of meson masses
\cite{pnjl:meson} as well as the validity and applicability of the
Taylor expansion in quark chemical potential used in LGT were
recently addressed within the PNJL model
\cite{sus:pnjl,pnjl:iso,RRW:phase}. In Ref.~\cite{pnjl:isov} the
model was extended to a system with finite isospin chemical
potential and pion condensation was studied.

In this  paper we will consider the fluctuations in various
channels at points in the phase diagram near the phase boundary. In
particular, we focus on the behavior of the net--quark number
fluctuations and the susceptibilities of the order
parameters\footnote{We will be somewhat lax, and refer to the
Polyakov loop and the quark condensate as order parameters,
although, due to the explicit breaking of the symmetries, neither
of them is strictly speaking an order parameter. However, as found
in LGT calculations, both are useful quantities for identifying the
location of the phase boundary. Thus, it seems that the explicit
breaking of both the chiral and center symmetry in QCD is in this
sense weak.}. The susceptibilities of the Polyakov loop and its
conjugate as well as the chiral condensate will be introduced and
analyzed. The relation between the chiral and deconfinement phase
transitions will be quantified through the susceptibilities. Our
calculations are performed within the mean field approximation and
in the PNJL model, including a non-local four--fermion interactions
to regulate the divergent momentum integrals.

The paper is organized as follows: In Section~\ref{sec:PNJL} the
PNJL model is formulated and the relevant thermodynamic potentials
are derived. In Section~\ref{sec:SCSB} the influence of the
Polyakov loop dynamics on various thermodynamical quantities is
discussed. In Section~\ref{sec:def_sus} and Section~\ref{sec:phase}
we introduce the susceptibilities of the order parameters and focus
on their properties. Concluding remarks and discussion are
presented in Section~\ref{sec:sum}.


\setcounter{equation}{0}
\section{Two-flavor NJL model with the Polyakov loop}
\label{sec:PNJL}

Various methods have been proposed to account for interactions with
the color gauge field in effective chiral
models~\cite{Gocksch,Meisinger,Fukushima,DLS,Megias,PNJL}. One of
these is the extension of the Nambu--Jona-Lasinio (NJL) Lagrangian
by coupling the quarks to a uniform temporal background gauge
field, which manifests itself entirely in the Polyakov loop
\cite{Meisinger,Fukushima,PNJL}. The PNJL Lagrangian for three
colors ($N_c = 3$) and two flavors ($N_f
= 2$) with non-local four-fermion
interactions is given by
\begin{eqnarray}
{\mathcal L}
&=& \bar{\psi} (i\Slash{D} - \hat{m})\psi
{}+ \bar{\psi}\hat{\mu}\gamma_0\psi
{}- {\mathcal U}(\Phi[A],\bar{\Phi}[A];T)
\nonumber\\
&&
{}+ \frac{G_S}{2}\left[\left(\bar{q}(x)q(x)\right)^2
{}+ \left(\bar{q}(x) i\gamma_5\vec{\tau}q(x)\right)^2 \right]\,,
\label{lag}
\end{eqnarray}
where $\hat{m} = \mbox{diag}(m_u, m_d)$ is the current quark mass,
$\hat{\mu} =
\mbox{diag}(\mu_u,\mu_d)$ is the quark chemical potential  and   $\vec{\tau}$ are the Pauli
matrices. We assume isospin symmetry and take $m_u
= m_d
\equiv m_0$ and $\mu_u =
\mu_d \equiv \mu$.

The Lagrangian is formulated with non-local interactions, which are
controlled by a form factor. This feature of the model is
implemented in order to deal with the ultraviolet singularities
that appear in the loop integrations. In coordinate space, the form
factor $F(x)$ for the non-local current-current interaction reads:
\begin{equation}
q(x) = \int d^4y F(x-y)\psi(y)\,.
\end{equation}
A possible choice for the regulator is in momentum space given
by~\cite{nonlocal}:
\begin{equation}
f^2(p) = \frac{1}{1 + (p/\Lambda)^{2\alpha}}\,,
\end{equation}
where $f(p)$ is the Fourier transform of the form factor $F(x)$ and
$p$ is the three--momentum.

The strength of the interaction among constituent quarks in
(\ref{lag}) is parameterized by the coupling constant $G_S$ which
carries the dimension of a length squared. In the pure NJL sector,
the model is controlled by four parameters: the coupling constant
$G_S$, the current quark mass $m_0$ and the constants $\alpha$ and
$\Lambda$, which characterize the range of the non-locality. These
parameters are determined in vacuum, for a given $\alpha$, by
requiring that the experimental values of the pion decay constant
$f_\pi
= 92.4$ MeV and  the pion mass $m_\pi = 135$ MeV as well as the
dynamical quark mass $M_{p=0}
= 335$ MeV are  reproduced. For $\alpha = 10$ the model
parameters are $\Lambda = 684.2$ MeV, $G_S\Lambda^2 = 4.66$, $m_0
= 4.46$ MeV. The corresponding value of the quark condensate is $\langle
\bar{\psi}\psi
\rangle^{1/3} = - 256.2$ MeV~\cite{nl:parameters}. The relevant
parameters of the model used in our calculations are summarized in
Table~\ref{table:njl}.

The interaction between the effective gluon field and the quarks is
in the PNJL Lagrangian implemented (\ref{lag}) by means of a
covariant derivative
\begin{equation}
D_\mu = \partial_\mu - iA_\mu\,, \quad \quad A_\mu =
\delta_{\mu 0}A^0\,,
\end{equation}
where we introduce the standard short-hand notation $A_\mu = g A_\mu^a
\frac{\lambda^a}{2}$. Here $g$ is the color SU(3)
gauge coupling constant and $\lambda^a$ are the Gell-Mann matrices.

The effective potential ${\mathcal U}$ of the gluon field in
(\ref{lag}) is expressed in terms of the traced Polyakov loop
$\Phi$  and its conjugate  $\bar{\Phi}$
\begin{equation}
\Phi = \frac{1}{N_c}\mbox{Tr}_c L\,, \qquad \bar{\Phi} = \frac{1}{N_c}\mbox{Tr}_c
L^\dagger\,, \label{phi}
\end{equation}
where $L$ is a matrix in color space related to the gauge field by
\begin{equation}
L(\vec{x}) = {\mathcal P}\exp\left[i\int_0^\beta d\tau
A_4(\vec{x},\tau)\right]\,,
\end{equation}
with ${\mathcal P}$ being the path (Euclidean time) ordering, and $\beta = 1/T$ with $A_4
= iA_0$.

In the heavy quark mass limit  QCD has the Z(3) center symmetry
which is spontaneously broken  in the high-temperature phase. The
thermal expectation value of the Polyakov loop $\langle \Phi
\rangle$ acts as an order parameter of the Z(3) symmetry.
Consequently, $\langle\Phi\rangle = 0$ at low temperatures in the
confined phase and $\langle \Phi \rangle \neq 0$ at high
temperatures corresponding to the deconfined phase. For the
SU$_c$(3) color gauge group the Polyakov loop matrix $L$ satisfies
$L L^\dagger = 1$, ${\mbox{det}}L = 1$  and can be written in
diagonal form
\begin{equation}
L = \mbox{diag}\left( e^{i\varphi}, e^{i\varphi^\prime}, e^{-i(\varphi + \varphi^\prime)}
\right)\,. \label{Lmatrix}
\end{equation}
In general $\Phi$ in Eq. (\ref{phi})    is not identical to
$\bar{\Phi}$.

The effective potential $\mathcal { U}(\Phi,\bar{\Phi})$ of the
gluon field is expressed in terms of the Polyakov loops so as to
preserve the $Z(3)$ symmetry of the pure gauge
theory~\cite{dumitru}. We adopt an   effective potential of the
following form~\cite{PNJL}~\footnote{The presence of quarks will
clearly affect the Polyakov loop dynamics. Thus, one may ask why
this is not reflected in the effective potential ${\mathcal U}$.
However, at least partly such effects are treated explicitly in the
PNJL model through the coupling to the quarks. Thus, in order to
minimize the double counting problem, we keep the $Z(3)$ symmetric
form of the effective potential, but consider the possibility that
the parameters, in particular $T_0$, may depend on the number of
dynamical flavors $N_f$, thus accounting for quark loop effects not
included in the mean-field approximation.}:
\begin{equation}
\frac{{\mathcal U}(\Phi,\bar{\Phi};T)}{T^4}
= - \frac{b_2(T)}{2}\bar{\Phi}\Phi
{}- \frac{b_3}{6}(\Phi^3 + \bar{\Phi}^3)
{}+ \frac{b_4}{4}(\bar{\Phi}\Phi)^2\,,
\label{eff_potential}
\end{equation}
with
\begin{equation}
b_2(T) = a_0 + a_1\left(\frac{T_0}{T}\right)
{}+ a_2\left(\frac{T_0}{T}\right)^2
{}+ a_3\left(\frac{T_0}{T}\right)^3\,.
\end{equation}
The coefficients $T_0$, $a_i$ and $b_i$ are fixed by requiring that
the equation of state obtained  in pure gauge theory on the lattice
is reproduced. In particular, at $T_0 = 270$ MeV the model
reproduces the first order deconfinement phase transition of the
pure gauge theory. The parameters are listed in
Table~\ref{table:pol}.

In the mean field approximation the Lagrangian~(\ref{lag})
is rewritten as
\begin{eqnarray}
{\mathcal L} &=& \bar{\psi}(i\Slash{D} +\hat{\mu}\gamma_0)\psi
{}-\int d^4y_1d^4y_2\bar{\psi}(y_1)M(y_1,y_2,x)\psi(y_2)
\nonumber\\
&& {}-
\frac{G_S}{2} \left(\langle\bar{q}(x)q(x)\rangle\right)^2
- {\mathcal U}(\Phi,\bar{\Phi};T)\,,\label{lagm}
\end{eqnarray}
where
\begin{eqnarray}
&&M(y_1,y_2,x)=m_0\delta^{(4)}(x-y_1)\delta^{(4)}(x-y_1)\nonumber\\
&&{}+F(x-y_1)F(x-y_2)\langle\bar{q}(x)q(x)\rangle\,.
\end{eqnarray}
In momentum space the dynamical quark mass is determined by the
current quark mass $m_0$,  the quark momentum distribution function
$f(p)$ and the chiral condensate $\langle\bar{q}q \rangle$
\begin{eqnarray}
&&
M_p = m_0 + \left( M - m_0 \right) f^2(p)\,,
\nonumber\\
&& M = m_0 - G_S \langle \bar{q}q \rangle\,,
\end{eqnarray}
where  $M$ denotes  the dynamical quark mass at $p=0$. In a uniform
system, the chiral condensate
\begin{equation}
\langle\bar{q}q\rangle = \int d^4p\,f(p)^2\,\langle\bar{\psi}(p)\psi(p)\rangle
\end{equation}
is independent of the position.

From the mean field Lagrangian (\ref{lagm})  one obtains the thermodynamic potential in
the following form
\begin{eqnarray}
&&
\Omega(M,\Phi,\bar{\Phi};T,\mu)
= {\mathcal U}(\Phi,\bar{\Phi};T)
{}+ \frac{(m_0 - M)^2}{2\,G_S}
\nonumber\\
&&
{}- 6 N_f \int\frac{d^3 p}{(2\pi)^3}
\left[ E_p - E_p(M_p=m_0) \right]
\nonumber\\
&&
{}- 2 N_f T \int\frac{d^3 p}{(2\pi)^3}\Bigl\{
\mbox{Tr}_c \ln \left[ 1 + L e^{-E^{\rm (+)}/T} \right]
\nonumber\\
&&\quad
{}+ \mbox{Tr}_c \ln \left[ 1 + L^\dagger e^{-E^{\rm (-)}/T} \right]
\Bigr\}\,,
\nonumber\\
\label{omega}
\end{eqnarray}
where we have introduced $E^{\rm (\pm)} = E_p \mp \mu$ for the
particle $(+)$ and anti-particle $(-)$ with $E_p =
\sqrt{|\vec{p}|^2 + M_p^2}$ being a quasiparticle energy. The third
term in the thermodynamic potential (\ref{omega}), which
corresponds to the vacuum contribution, is renormalized by
subtracting the term $E_p(M_p=m_0)$ under the momentum integral.
This removes the divergence from and introduces an irrelevant
constant shift of $\Omega$.

Furthermore, by taking the trace in color space, one obtains the
final expression for the thermodynamic potential
\begin{eqnarray}
&&
\Omega(M,\Phi,\bar{\Phi};T,\mu)
= {\mathcal U}(\Phi,\bar{\Phi};T)
{}+ \frac{(m_0 - M)^2}{2\,G_S}
\nonumber\\
&&
{}- 6 N_f \int\frac{d^3 p}{(2\pi)^3}
\left[ E_p - E_p(M_p=m_0) \right]
\nonumber\\
&&
{}- 2 N_f T \int\frac{d^3 p}{(2\pi)^3}\Bigl\{
\ln\left[ g^{\rm (+)}(M,\Phi,\bar{\Phi};T,\mu;p)\right]
\nonumber\\
&&\quad
{}+ \ln\left[ g^{\rm (-)}(M,\Phi,\bar{\Phi};T,\mu;p)\right]
\Bigr\}\,,
\nonumber\\
\label{omega2}
\end{eqnarray}
where
\begin{eqnarray}
&&
g^{\rm (+)}(M,\Phi,\bar{\Phi};T,\mu;p)
\nonumber\\
&&
= 1 + 3\left( \Phi + \bar{\Phi}e^{-E^{\rm (+)}/T} \right)
e^{-E^{\rm (+)}/T} + e^{-3E^{\rm (+)}/T}\,,
\nonumber\\
&&
g^{\rm (-)}(M,\Phi,\bar{\Phi};T,\mu;p)
\nonumber\\
&&
= 1 + 3\left( \bar{\Phi} + \Phi e^{-E^{\rm (-)}/T} \right)
e^{-E^{\rm (-)}/T} + e^{-3E^{\rm (-)}/T}\,.
\nonumber\\
\label{log-arg}
\end{eqnarray}
Under the transformation $\mu\to -\mu$, the role of quarks and
antiquarks is exchanged. Inspection of (\ref{log-arg}) shows that
in the PNJL model the charge conjugation transformation also
exchanges the role of the Polyakov loop and its conjugate. This is
reflected in the relation
$g^{(+)}(M,\Phi,\bar{\Phi};T,\mu;p)=g^{(-)}(M,\bar{\Phi},\Phi;T,-\mu;p)$.

Although the Polyakov loop is complex, the thermodynamic potential
is real. Due to the symmetry in color space, the imaginary part of
$\Omega $ vanishes after performing the functional integral [see
Appendix~\ref{app:basis}].

An interesting feature of the PNJL model described by the thermodynamic potential
(\ref{omega2}) is the qualitative behavior in the low temperature phase. In the limit of
$\Phi,\bar{\Phi}\to 0$, which is expected at low temperatures, the contribution of one-
and two-quark states to $g^{(\pm)}$ are suppressed and only the three--quark term
$\sim\exp(-3E^{(\pm)}/T)$ survives. In this sense the PNJL model mimics the confinement
of quarks within three-quark states. The suppression of quark degrees of freedom at low
temperatures is, on a qualitative level, similar to confinement in QCD thermodynamics.
Thus, the PNJL model is better suited for describing the low temperature QCD phase than
the standard NJL model, where the constituent quarks are abundant also at low
temperatures. However, at least in the mean-field approximation, the model only has the
three-quark states, but there are no one- and two-quark states, which also play an
important role at low temperatures.

In the mean field approximation  the dynamical quark mass $M$ and
the expectation values of the Polyakov loop $\Phi$ and $\bar{\Phi}$
are obtained from the stationarity conditions
\begin{equation}
\frac{\partial\Omega}{\partial M} = \frac{\partial\Omega}{\partial \Phi} =
\frac{\partial\Omega}{\partial \bar{\Phi}} = 0\,,
\end{equation}
obtained by extremizing the thermodynamic potential with respect to
$M$, $\Phi$ and $\bar{\Phi}$.
\begin{widetext}
These conditions yield the following set of coupled gap equations:
\begin{eqnarray}
&&
M = m_0 + 6\,G_S N_f \int\frac{d^3 p}{(2\pi)^3}
\frac{M_p f^2(p)}{E_p}\Biggl[ 1
{}- \frac{(\Phi + 2\bar{\Phi}e^{-E^{\rm (+)}/T}
{}+ e^{-2E^{\rm (+)}/T})e^{-E^{\rm (+)}/T}}
{g^{\rm (+)}(M,\Phi,\bar{\Phi};T,\mu;p)}
\nonumber\\
&&
\hspace*{6.1cm}
{}- \frac{(\bar{\Phi} + 2\Phi e^{-E^{\rm (-)}/T}
{}+ e^{-2E^{\rm (-)}/T})e^{-E^{\rm (-)}/T}}
{g^{\rm (-)}(M,\Phi,\bar{\Phi};T,\mu;p)}
\Biggr]\,,
\label{gapeq:M}
\\
&&
b_2(T)\bar{\Phi} + b_3 \Phi^2 - b_4 (\bar{\Phi}\Phi)\bar{\Phi}
= - \frac{12 N_f}{T^3}\int\frac{d^3 p}{(2\pi)^3}
\Biggl[
\frac{e^{-E^{\rm (+)}/T}}{g^{\rm (+)}(M,\Phi,\bar{\Phi};T,\mu;p)}
{}+
\frac{e^{-2E^{\rm (-)}/T}}{g^{\rm (-)}(M,\Phi,\bar{\Phi};T,\mu;p)}
\Biggr]\,,
\label{gapeq:Phi}
\\
&&
b_2(T)\Phi + b_3 \bar{\Phi}^2 - b_4 \Phi(\bar{\Phi}\Phi)
= - \frac{12 N_f}{T^3}\int\frac{d^3 p}{(2\pi)^3}
\Biggl[
\frac{e^{-2E^{\rm (+)}/T}}{g^{\rm (+)}(M,\Phi,\bar{\Phi};T,\mu;p)}
{}+
\frac{e^{-E^{\rm (-)}/T}}{g^{\rm (-)}(M,\Phi,\bar{\Phi};T,\mu;p)}
\Biggr]\,.
\label{gapeq:Phi-bar}
\end{eqnarray}
\end{widetext}
As noted above, the role of $\Phi$ and $\bar{\Phi}$ are exchanged
under charge conjugation ($\mu\to -\mu$). This symmetry is
reflected in the relation between the gap
equations~(\ref{gapeq:Phi}) and (\ref{gapeq:Phi-bar}) under the
charge conjugation transformation. In particular, this relation
implies that $\Phi=\bar{\Phi}$ for $\mu= 0$. Furthermore, we note
that for $\Phi=\bar{\Phi} = 1$, Eq.~(\ref{gapeq:M}) is reduced to
the gap equation of the standard NJL model without any coupling to
the color SU$_c$(3) gauge field.

\setcounter{equation}{0}
\section{Thermodynamic quantities at finite quark chemical potentials}
\label{sec:SCSB}

\begin{figure}
\begin{center}
\includegraphics[width=8cm]{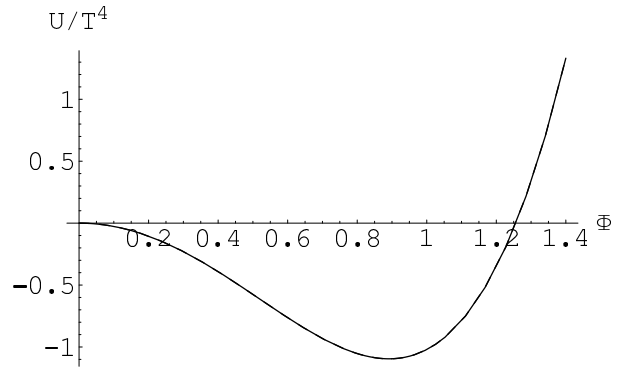}
\caption{\label{fig:poly} The Polyakov loop effective potential ${\mathcal U}/T^4$ as a
function of $\Phi$ at $T=500$ MeV in the chiral limit. }
\end{center}
\end{figure}

\begin{figure}
\begin{center}
\includegraphics[width=8cm]{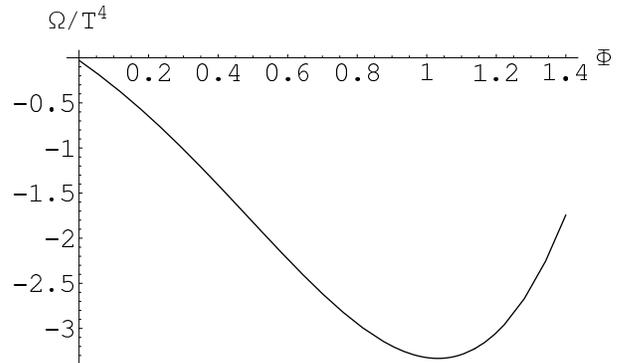}
\caption{\label{fig:omega} The thermodynamic potential $\Omega/T^4$ as a function of
$\Phi$ at $T=500$ MeV in the chiral limit. }
\end{center}
\end{figure}

The thermodynamics  of the PNJL model in the  mean-field
approximation is characterized by the potential $\Omega(\Phi ,\bar
{\Phi}, M)$ introduced in Eq. (\ref{omega2}). In
Figs.~\ref{fig:poly} and~\ref{fig:omega} we show the effective
potential of the Polyakov loop $\mathcal{U}$ as well as the PNJL
thermodynamic potential $\Omega$ in the chiral limit, at a high
temperatures where chiral symmetry is restored and the dynamical
quark mass $M$ vanishes. Consequently, only the Polyakov loop and
its conjugate are the relevant classical fields in the problem.

For vanishing quark chemical potential, $\Phi$ and $\bar
\Phi$ are equivalent. Hence, the thermodynamic potential is characterized by only
one variable $\Phi$ that for a given temperature is determined by
the gap equation (\ref{gapeq:Phi}). In Fig. ~\ref{fig:poly} we show
the $\Phi$ dependence of $\mathcal U$. This  potential exhibits the
expected structure in the $Z(3)$ symmetry broken phase with a
minimum at a finite value of $\Phi$ below unity.

The influence of quarks on the PNJL model thermodynamics is
illustrated in Fig.~\ref{fig:omega}, where the $\Phi$--dependence
of $\Omega$ is shown for the same temperature, $T=0.5 $ GeV. A
comparison of Figs.~\ref{fig:poly} and~\ref{fig:omega} clearly
shows that the interactions of the effective gluon field with the
quarks leads to a shift of the minimum to larger values of $\Phi$.
At high temperatures the minimum corresponds to $\Phi > 1$.


\begin{figure}
\begin{center}
\includegraphics[width=8cm]{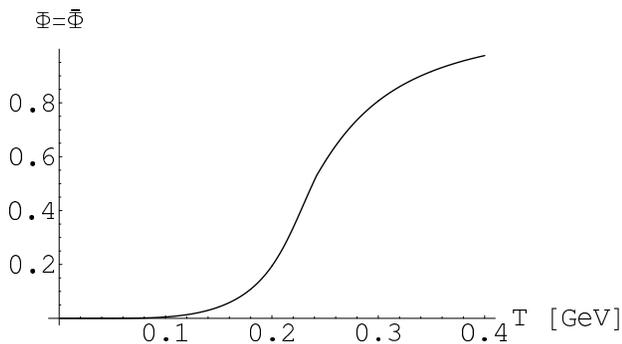}
\caption{\label{fig:soln_Phi}
Expectation value of the traced Polyakov loop $\Phi$
in the chiral limit as a function of temperature $T$
for vanishing quark chemical potential.
}
\end{center}
\end{figure}

\begin{figure}
\begin{center}
\includegraphics[width=8cm]{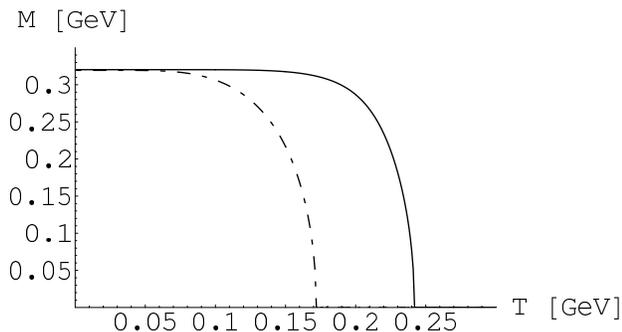}
\caption{\label{fig:soln_M} The dynamical quark mass $M$ in the chiral limit as a
function of temperature $T$ for a vanishing quark chemical potential. The dash-dotted
line denotes the result obtained in the NJL model. }
\end{center}
\end{figure}

The dynamical quark mass $M$ and the traced Polyakov loop $\Phi$
are the order parameters of the chiral and $Z(3)$ symmetries
respectively. Thus, the $T$ and $\mu$ dependence of these order
parameters is used to identify the phase boundaries of the model.
In Figs.~\ref{fig:soln_Phi} and~\ref{fig:soln_M} we show $\Phi$ and
$M$ at $\mu = 0$, as functions of $T$ in the chiral limit. The
phase change of the model is clearly indicated by rapid changes of
the order parameters. The  Polyakov loop  potential  $\mathcal U$
introduced in the PNJL Lagrangian preserves the invariance under
the $Z(3)$ symmetry. However, due to the interactions with the
quarks this symmetry is explicitly broken in the model. Thus, the
Polyakov loop is not an order parameter in the strict sense and the
transition is a rapid cross over. This is a remnant of the
"deconfinement" phase transition of the pure Polyakov loop model.
On the other hand, in the chiral limit the chiral transition is a
true second order phase transition, with an order parameter $M$,
which is strictly zero for temperatures above $T\simeq 242$ MeV.

In Fig.~\ref{fig:soln_M} the  dynamical quark mass  $M$ obtained in
the  PNJL model is also compared with that of the NJL
model~\footnote{In the NJL model, without Polyakov loops, we use
the local version of the cutoff, i.e., a sharp cutoff. The cutoff
is not implemented in the thermal part, which is anyway
convergent.}. It is clear that the coupling of the quarks to the
effective gluon field shifts the chiral phase transition
temperature to the higher value. This is a consequence of
attractive interactions; in the presence of the Polyakov loop there
is stronger "binding" of the constituent quarks in the chirally
broken phase.

The non-locality  of the quark interactions introduced by the form
factor $F(x)$ also affects the critical temperature of the chiral
phase transition. Suppressing the form factor in the PNJL
Lagrangian results in a lower transition temperature. With the
parameters used in our actual calculations, the reduction of $T_c$
is on the order of $15$ MeV, if in the thermal part of the  PNJL
model the momentum cutoff is not implemented.

\begin{figure*}
\begin{center}
\includegraphics[width=8cm]{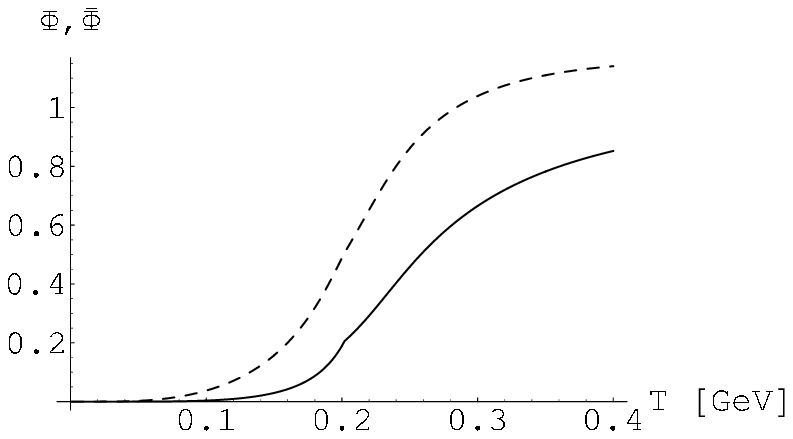}
\includegraphics[width=8cm]{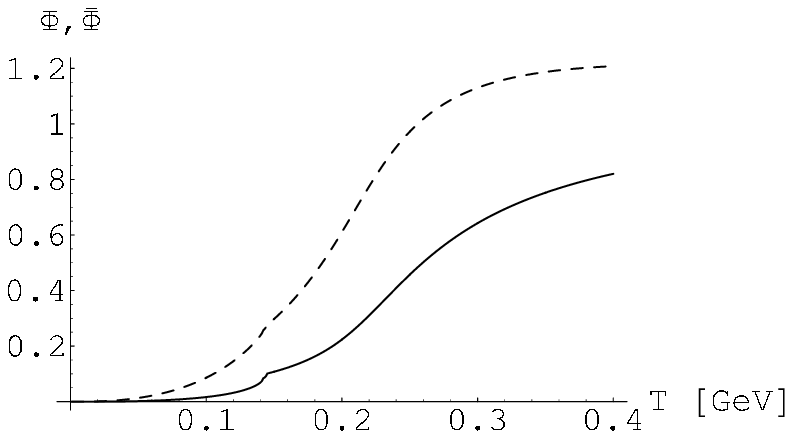}
\caption{\label{fig:soln-mu_Phi} Expectation values of the traced Polyakov loop $\Phi$
(solid) and  $\bar{\Phi}$ (dashed) in the chiral limit as a function of temperature $T$
for finite quark chemical potentials. The lines in the left figure show $\Phi$ and
$\bar{\Phi}$ at $\mu=200$ MeV and the chiral phase transition is of second order. The
right figure is the results obtained at $\mu=270$ MeV, which correspond to the first
order chiral phase transition.
 }
\end{center}
\end{figure*}

\begin{figure}
\begin{center}
\includegraphics[width=8cm]{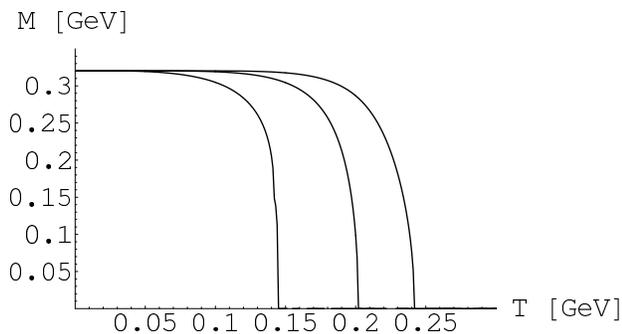}
\caption{\label{fig:soln-mu_M} The dynamical quark mass $M$ in the chiral limit as a
function of temperature $T$ for finite quark chemical potentials $\mu = 0$, $200$ MeV,
$270$ MeV from right to left.
 }
\end{center}
\end{figure}

\begin{figure}
\begin{center}
\includegraphics[width=8cm]{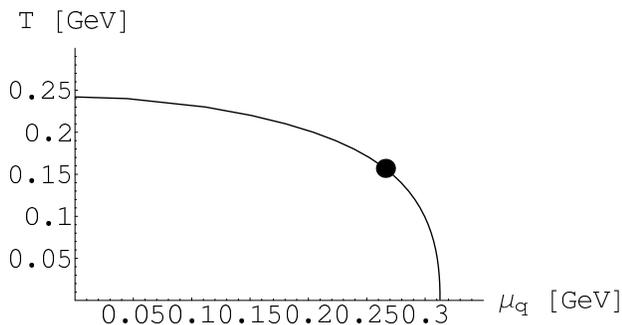}
\caption{\label{fig:phase_ch} The phase diagram of the PNJL model in the chiral limit.
The dot indicates the location of the tricritical point, located at
$(T_{TCP}=157, \mu_{TCP}=266)$ MeV. }
\end{center}
\end{figure}

The introduction of a finite quark chemical potential is expected
to modify the behavior of the order parameters. The $\mu$
dependence of $M$, $\Phi$ and $\bar{\Phi}$ are shown  in
Figs.~\ref{fig:soln-mu_Phi} and~\ref{fig:soln-mu_M}. There is a
clear shift of the chiral transition to lower temperatures with
increasing $\mu$ as is seen in Fig.~\ref{fig:soln-mu_M}. At a
non-zero quark chemical potential the charge conjugation symmetry
is broken, which leads to a splitting  between $\Phi$ and
$\bar\Phi$. As seen in Fig.~\ref{fig:soln-mu_Phi} the Polyakov loop
$\Phi$ is decreasing whereas $\bar\Phi$ is increasing with $\mu$.
This is expected due to the relation of the Polyakov loop and its
conjugate to the free energy of a quark and an antiquark
respectively \cite{mclerran}. In a system with more quarks than
antiquarks it is relatively easy to screen a static antiquark by a
quark, forming a virtual $q\bar q$ state, whereas a static quark
can only be screened by a diquark, thereby forming a colorless
three-quark state.

At finite quark chemical potential, the order of the chiral phase
transition can change. This is illustrated in
Fig.~\ref{fig:phase_ch} where  the phase diagram of the PNJL model
is shown. At low temperatures and large $\mu$ the transition is
first order. The first order transition terminates at the
tricritical point (TCP). With the actual value of the model
parameters, summarized in Table.~\ref{table:Tc_adj}, the TCP
appears at $(T_{TCP}=157,\mu_{TCP}=266)$ MeV. Beyond the TCP, at
smaller chemical potentials, the transition is second order. This
is consistent with the gross structure of the phase diagram,
expected for QCD according to universality arguments
\cite{pisarski}.

We explore the influence of the Polyakov loop on the
thermodynamics, by considering observables that are related to the
conservation of the net--quark number such as the quark number
density $n_q(T,\mu )$ and the corresponding susceptibility $\chi_q(
T,\mu )$. Both observables are obtained as derivatives of the
thermodynamic potential $\Omega$ with respect to $\mu$. The quark
number density is obtained from
\begin{eqnarray}
n_q = - \frac{\partial\Omega}{\partial\mu}\,.
\end{eqnarray}
With the thermodynamic potential of the PNJL model (\ref{omega2})
one finds
\begin{eqnarray}
n_q
&=& 6 N_f \int\frac{d^3 p}{(2\pi)^3}
\Biggl[ \frac{e^{-E^{\rm (+)}/T}}{g^{\rm (+)}}
\nonumber\\
&&\times
\left( \Phi + 2\bar{\Phi} e^{-E^{\rm (+)}/T}
{}+ e^{-2E^{\rm (+)}/T}
\right)
\nonumber\\
&& {}- \frac{e^{-E^{\rm (-)}/T}}{g^{\rm (-)}} \left( \bar{\Phi} + 2\Phi e^{-E^{\rm
(-)}/T} {}+ e^{-2E^{\rm (-)}/T} \right) \Biggr]\,.
\nonumber\\
\label{n_q}
\end{eqnarray}

\begin{figure}
\begin{center}
\includegraphics[width=8cm]{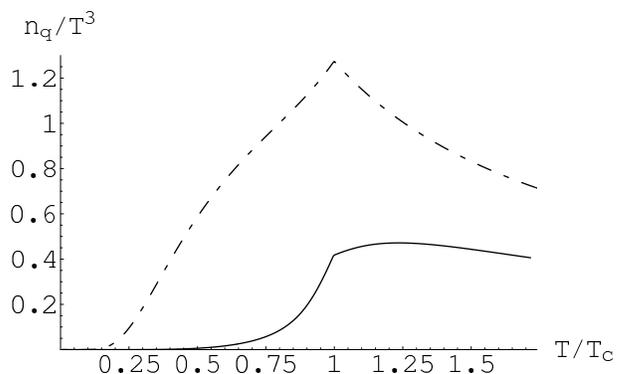}
\caption{\label{fig:nq-ff} The net--quark number density $n_q/T^3$ in the chiral limit as
a function of temperature $T$ at $\mu
= 100$ MeV. The dash-dotted line denotes the
result obtained in the NJL model.
 }
\end{center}
\end{figure}

\begin{figure}
\begin{center}
\includegraphics[width=8cm]{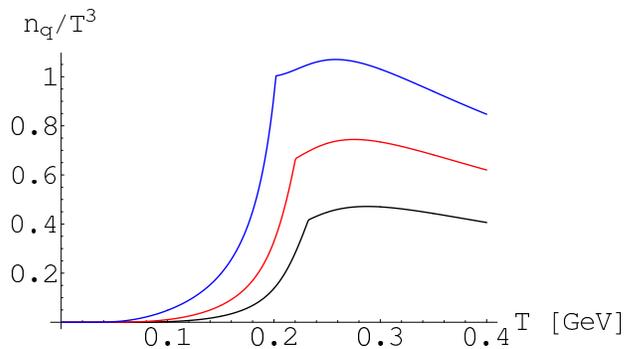}
\caption{\label{fig:nq-mu} The net--quark number density $n_q/T^3$ in the chiral limit as
a function of temperature $T$. The lines, starting from the lowest
one, correspond $\mu = 100, 150, 200$ MeV.
 }
\end{center}
\end{figure}

The temperature dependence of the net--quark number density
obtained with Eq. (\ref{n_q}) is shown in Figs.~\ref{fig:nq-ff}
and~\ref{fig:nq-mu} for various values of $\mu$. The PNJL model
results are also compared with those of the NJL model. Clearly
there is a substantial change in $n_q$ when the Polyakov loop
dynamics is included.

In the region above the chiral phase transition  the standard NJL model shows a
relatively strong decrease of $n_q/T^3$  with increasing temperature. This behavior
exactly reproduces the temperature dependence of a non-interacting gas of massless
quarks, $n_q=N_f(\mu^3/\pi^2+\mu T^2)$. Thus, the leading term in $n_q/T^3$ is
proportional to $\mu/T$. Furthermore, the quark density increases as the critical
temperature $T_c$ is approached from below. This is due to the decrease of the effective
quark mass as the chiral symmetry is restored. The PNJL model shows a substantially
different temperature dependence of the quark density. First, there is a suppression of
the one- and two-quark contributions to the density below $T_c$, due to the interactions
with the effective gluon field. Hence, the leading contribution to the net quark density
is due to the three--quark  states. Clearly this leads to a strong suppression of the
quark density below $T_c$. Above $T_c$, the suppression is much less effective. However,
because $\Phi$ is still less than unity, a suppression compared to the quark density of a
free gas at the same temperature and chemical potential remains.

The quark number susceptibility  measures the  response of the
quark number density to changes in the quark chemical potential.
This observable is of particular interest for exploring tricritical
point. This is because $\chi_q$ is expected to diverge at TCP. This
divergence is a remnant of the diverging fluctuations of the
scalar-isoscalar sigma field ~\cite{hatta,fujii,stephanov}. The
net--quark number susceptibility is defined by
\begin{eqnarray}
\chi_q = \frac{\partial n_q}{\partial\mu}\,. \label{chi}
\end{eqnarray}

In the PNJL model  the dynamical quark mass $M$ and the Polyakov
loops $\Phi,\bar{\Phi}$ implicitly depend on $\mu$. Thus, besides
an explicit  $\mu$-dependent  contribution through Eq. (\ref{chi})
there are also terms  proportional to $\mu$--derivatives of the
effective condensates. In the PNJL model one finds
\begin{eqnarray}
\chi_q
&=& \chi_q^{(0)}
{}+ T^2 A_M^{(\mu)}
\frac{\partial M}{\partial\mu}
{}+ T^3 A_\Phi^{(\mu)}
\frac{\partial\Phi}{\partial\mu}
{}+ T^3 A_{\bar{\Phi}}^{(\mu)}
\frac{\partial\bar{\Phi}}{\partial\mu}
\nonumber\\
&\equiv&
\chi_q^{(0)} + \chi_q^{(M)} + \chi_q^{(\Phi)}
{}+ \chi_q^{(\bar{\Phi})}\,,
\label{chiq}
\end{eqnarray}
where $\chi_q^{(0)}$ corresponds to the 0-th order contribution
given by
\begin{eqnarray}
\chi_q^{(0)}
&=&
\frac{6 N_f}{T} \int\frac{d^3 p}{(2\pi)^3}
\Biggl[
\frac{e^{-E^{(+)}/T}}{(g^{(+)})^2}\Bigl\{
\Phi + 4\bar{\Phi}e^{-E^{(+)}/T}
\nonumber\\
&&
{}+ 3 \left( 1 + \bar{\Phi}\Phi \right) e^{-2E^{(+)}/T}
{}+ 4\Phi e^{-3E^{(+)}/T}
\nonumber\\
&&
{}+ \bar{\Phi}e^{-4E^{(+)}/T}
\Bigr\}
{}+ \left( \bar{\Phi},\Phi;-\mu \right)
\Biggr]\,.
\end{eqnarray}
The functions $A^{(\mu)}(M,\Phi,\bar{\Phi};T,\mu)$ and the
$\mu$-derivatives of the condensates  are introduced  in
Appendix~\ref{app:der}.

\begin{figure}
\begin{center}
\includegraphics[width=8cm]{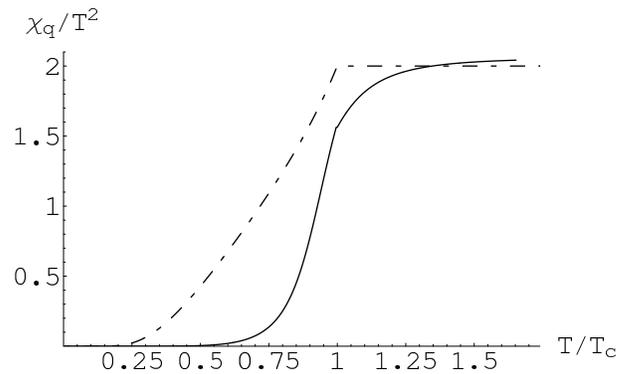}
\caption{\label{fig:qsus-ff} The quark number susceptibility $\chi_q/T^2$ in the chiral
limit as a function of temperature $T$, at $\mu = 0$. The
dash-dotted line denotes the result obtained in the NJL model.}
\end{center}
\end{figure}

\begin{figure}
\begin{center}
\includegraphics[width=8cm]{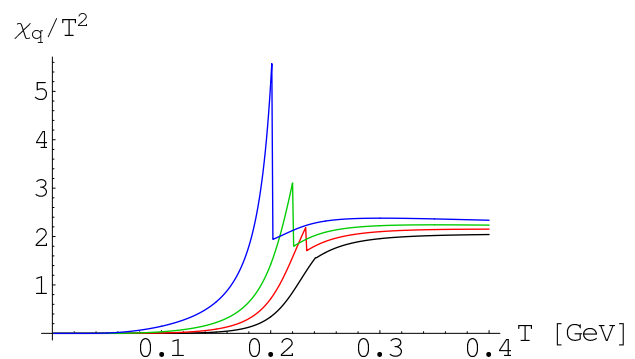}
\caption{\label{fig:qsus-mu} The quark number susceptibility $\chi_q/T^2$ in the chiral
limit as a function of temperature $T$. The lines correspond to the results in the PNJL
model at $\mu = 0, 100, 150, 200$ MeV from below. }
\end{center}
\end{figure}

The quark number susceptibility at vanishing and at finite $\mu$ is
shown as a function of the temperature in  Figs.~\ref{fig:qsus-ff}
and~\ref{fig:qsus-mu} for the PNJL and NJL models.
At high temperature $\chi_q$ approaches the ideal gas limit,
$\chi_q/T^2 \simeq 2$. In the chiral limit, the quark number
susceptibility has a discontinuity at the chiral phase transition
at finite $\mu$, as found in Landau-Ginzburg theory~\cite{hatta}
(see also~\cite{our,SFR1} for a detailed discussion of the quark number
susceptibility in the NJL model).

The effect of quark-gluon interactions on the quark number
susceptibility, shown in Fig.~\ref{fig:qsus-ff}, is similar to that
discussed in the context of the net--quark density. There is a
reduction of the quark fluctuations below $T_c$ in the PNJL model
relative to the fluctuations  obtained in NJL calculations. The
PNJL results for the $T$ and $\mu$ dependence of $\chi_q$ and $n_q$
are in good agreement with recent LGT calculations of these
quantities~\cite{lattice:sus2}. Thus, in contrast to the NJL model,
the PNJL model provides a quantitative description of QCD
thermodynamics near the phase transition.

\begin{figure*}
\begin{center}
\includegraphics[width=8cm]{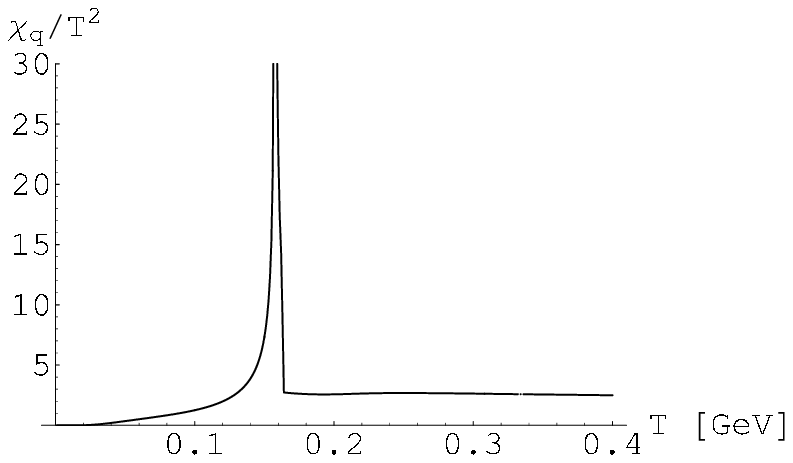}
\includegraphics[width=8cm]{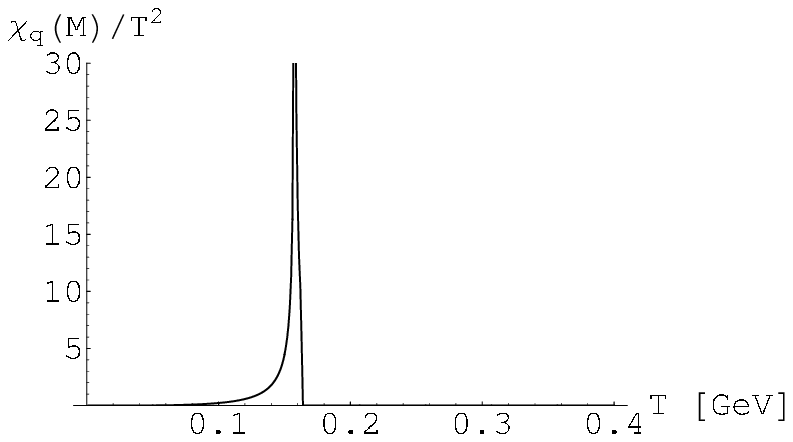}
\\
\includegraphics[width=8cm]{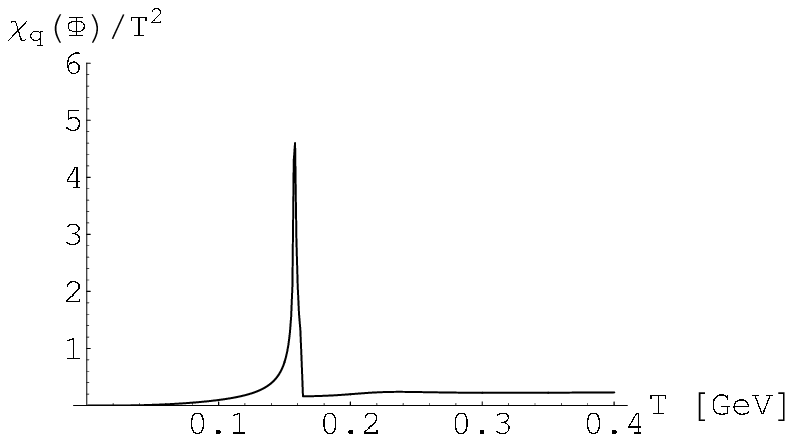}
\includegraphics[width=8cm]{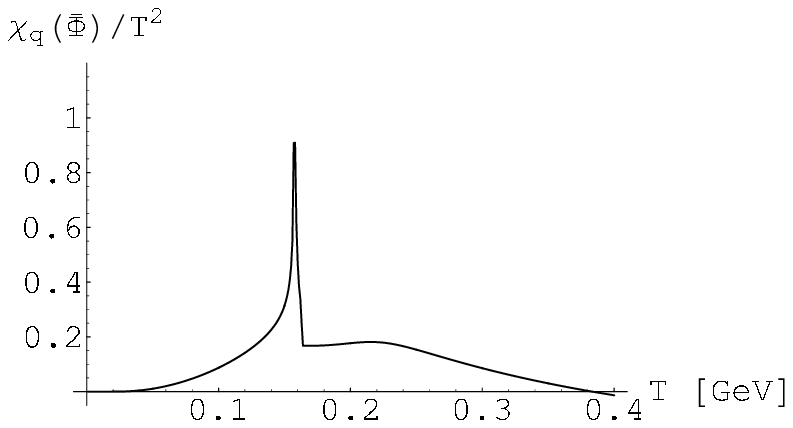}
\caption{\label{fig:qsus-tcp} The quark number susceptibility $\chi_q/T^2$ in the chiral
limit as a function of temperature $T$ at the TCP $\mu = 266$ MeV.
The components $\chi_q^{(M)}, \chi_q^{(\Phi)}$ and
$\chi_q^{(\bar{\Phi})}$ as a function of $T$ at the TCP are defined
in Eq.~(\ref{chiq}). }
\end{center}
\end{figure*}

At the TCP the quark number susceptibility  diverges as shown in the left panel of
Fig.~\ref{fig:qsus-tcp}.
We now analyze the different contributions in Eq.~(\ref{chiq}). The
right panel of Fig.~\ref{fig:qsus-tcp} clearly shows that the
singular behavior of $\chi_q$ is caused by the divergence of
$\chi_q^{(M)}$. Both $\chi_q^{(\Phi)}$ and $\chi_q^{(\bar{\Phi})}$
have sharp peak structures at $T_c$, but, in contrast to
$\chi_q^{(M)}$, they remain finite .

In order to understand the relation between the singular
contributions to the susceptibilities, we consider an effective
Lagrangian of the Landau type~\cite{LL}
\begin{equation}
\Omega(T,\mu) \simeq \Omega_0(T,\mu) + m M + \frac{1}{2}a(T,\mu)M^2
{}+ \frac{1}{4}b(T,\mu)M^4\,,
\label{Landau}
\end{equation}
where $a(T,\mu)$ and $b(T,\mu)$ are temperature dependent
coefficients and $m$ is a source term, which breaks the symmetry
explicitly. At the end of the calculation we consider the ``chiral
limit'', and set $m=0$. The equilibrium value of the order
parameter $M$ is determined by the gap equation $\partial
\Omega/\partial M=0$. In the broken symmetry phase $M=\sqrt{-a/b}$,
while in the symmetric phase $M=0$. The model exhibits a second
order transition at $a=0$ for $b>0$, and a tricritical point at
$a=b=0$ [see e.g., Ref.~\cite{hatta}]. The chiral susceptibility
\begin{equation}
\chi_{mm} =  \frac{\partial M}{\partial m} = \frac{1}{a(T,\mu) + 3 M^2 b(T,\mu)}\,,
\end{equation}
diverges along the second order transition line as well as at the
TCP in the broken phase.

On the other hand, the singular part of the quark number
susceptibility is given by
\begin{equation}
\chi_q^{(M)}=-\frac{\partial a}{\partial \mu}M\frac{\partial M}{\partial \mu}\,.
\end{equation}
By using the gap equation one finds
\begin{equation}
\frac{\partial M}{\partial \mu}\sim M\chi_{mm}\,,
\end{equation}
which implies
\begin{equation}
\chi_q^{(M)}\sim M^2\chi_{mm} \sim \frac{1}{b(T,\mu)}\,,
\end{equation}
where in the final step we used the gap equation to eliminate $M$.
Thus, the quark number susceptibility diverges only at the TCP. The
mixing of the chiral and quark number susceptibilities can also be
interpreted as a consequence of $\sigma$--$\omega$ mixing at finite
baryon density ~\cite{hatta,fujii,stephanov,SFR1}.


\setcounter{equation}{0}
\section{Chiral and Polyakov loop susceptibilities}
\label{sec:def_sus}

In the PNJL model the constituent quarks and the Polyakov loops are
effective fields related with the order parameters for the chiral
and $Z(3)$ symmetry breaking. In LGT the susceptibilities
associated with these fields show clear signals of the phase
transitions. In this section we present the calculational scheme
for computing the susceptibilities and discuss their relation to
the spontaneous breaking of the chiral and $Z(3)$ symmetries.

Consider first the generating functional of a scalar field theory
\begin{equation}
Z[J] = e^{-iW[J]}
= \int{\mathcal D}\phi \exp
\left[ i\int d^4x \left( {\mathcal L}[\phi]
{}+ J\phi \right)\right]\,,
\end{equation}
where $J$ is an external source. The second functional derivative
of $W[J]$ with respect to $J$ yields the correlation function:
\begin{equation}
\frac{\delta^2 W[J]}{\delta J(x)\delta J(y)}
\Big|_{J = 0}
= \langle \phi(x)\phi(y) \rangle - \langle \phi(x) \rangle
\langle \phi(y) \rangle\,.
\label{exp}
\end{equation}
The effective action $\Gamma[\phi]$ is introduced through the
Legendre transformation
\begin{equation}
\Gamma [\phi] = - W[J] - \int d^4 x J(x)\phi(x)\,.
\end{equation}
The correlation function introduced in the  Eq.~(\ref{exp}) can now
be expressed by means of
\begin{equation}
\frac{\delta^2 W[J]}{\delta J(x)\delta J(y)}\Big{|}_{J=0}
 = \left( \frac{\delta^2 \Gamma[\phi]}
{\delta \phi(x)\delta \phi(y)}\right)^{-1}\,. \label{legendre}
\end{equation}
Thus, by using this identity, one can compute the susceptibility
also from the effective potential, by taking derivatives with
respect to the classical field $\phi$.

In the case of the PNJL model there are three different classical
fields $\vec \phi=(M,\Phi,\bar {\Phi})$ that correspond to the
order parameters; the dynamical quark mass, the Polyakov loop and
its complex conjugate. Consequently, in order to compute the
corresponding susceptibilities from the effective potential, the
relation (\ref{legendre}) must be generalized. We use the chain
rule,
\begin{equation}\label{chain}
\delta_{ij}
=\frac{\delta \phi_i}{\delta \phi_j}
=\frac{\delta}{\delta \phi_j}\frac{\delta W}{\delta J_i}
=\frac{\delta J_k}{\delta \phi_j}\frac{\delta^2 W}{\delta J_k\delta J_i}
=\frac{\delta^2 \Gamma}{\delta \phi_j\delta \phi_k}
 \frac{\delta^2 W}{\delta J_k\delta J_i}\,,
\end{equation}
where $\vec J=(J_M,J_\Phi,J_{\bar{\Phi}})$ is a vector composed of
the source fields corresponding to the order parameters in
$\vec{\phi}$.

It follows from Eq.~(\ref{chain}) that the susceptibilities are
obtained by inverting the matrix composed of the second derivatives
of the effective action with respect to classical fields. Following
Ref.~\cite{Fukushima}, we introduce a dimensionless matrix
\begin{equation}
\hat{C} =
\left(\begin{array}{ccc}
C_{mm} & C_{ml} & C_{m\bar{l}} \\
C_{lm} & C_{ll} & C_{l\bar{l}} \\
C_{\bar{l}m} & C_{\bar{l}l} & C_{\bar{l}\bar{l}}
\end{array}\right)\,,
\end{equation}
with the components
\begin{eqnarray}
&&
C_{mm}
= \frac{1}{T\Lambda}\frac{\partial^2\Omega}{\partial M^2}\,,
\quad
C_{ll}
= \frac{1}{T\Lambda^3}\frac{\partial^2\Omega}{\partial\Phi^2}\,,
\nonumber\\
&&
C_{\bar{l}\bar{l}}
= \frac{1}{T\Lambda^3}
\frac{\partial^2\Omega}{\partial\bar{\Phi}^2}\,,
\quad
C_{l\bar{l}} = C_{\bar{l}l}
= \frac{1}{T\Lambda^3}
\frac{\partial^2\Omega}{\partial\Phi \partial\bar{\Phi}}\,,
\nonumber\\
&&
C_{ml} = C_{lm}
= \frac{1}{T\Lambda^2}
\frac{\partial^2\Omega}{\partial M \partial\Phi}\,,
\nonumber\\
&&
C_{m\bar{l}} = C_{\bar{l}m}
= \frac{1}{T\Lambda^2}
\frac{\partial^2\Omega}{\partial M \partial\bar{\Phi}}\,.
\end{eqnarray}
Through Eq.~(\ref{chain}) a set of susceptibilities is defined by
\begin{equation}
\hat{\chi} =
\left(\begin{array}{ccc}
\chi_{mm} & \chi_{ml} & \chi_{m\bar{l}} \\
\chi_{lm} & \chi_{ll} & \chi_{l\bar{l}} \\
\chi_{\bar{l}m} & \chi_{\bar{l}l} & \chi_{\bar{l}\bar{l}}
\end{array}\right)\,,
\label{chi-matrix}
\end{equation}
where $\chi_{ij}$ is given by the inverse of $\hat{C}$,
\begin{equation}
\chi_{ij} = \left( \hat{C}^{-1} \right)_{ij}
\quad
i,j = \{ m, l, \bar{l} \}\,.
\end{equation}
Here $\chi_{mm}$ is the chiral susceptibility, as in the previous
section, while $\chi_{ll}$ and $\chi_{\bar{l}\bar{l}}$ are the
diagonal Polyakov loop susceptibilities (see below). The
off-diagonal terms correspond to mixed susceptibilities.

In the pure gluon sector, the susceptibilities  are related to the
fluctuations of the $\Phi$ and $\bar{\Phi}$ fields. Under the
$Z(3)$ transformation $\bar{\Phi}\Phi,
\Phi^3, \bar{\Phi}^3$ and combinations thereof are invariant. Thus, the  off-diagonal
susceptibility $\chi_{l\bar{l}}$ is $Z(3)$ invariant, while the
diagonal pieces $\chi_{ll,\bar{l}\bar{l}}$ are not:
\begin{eqnarray}
\chi_{l\bar{l}}
&=&
\langle \bar{\Phi}\Phi \rangle - \langle \bar{\Phi} \rangle \langle \Phi \rangle\,,
\nonumber\\
\chi_{ll}
&=&
\langle \Phi^2 \rangle - \langle \Phi \rangle^2\,,
\nonumber\\
\chi_{\bar{l}\bar{l}}
&=&
\langle \bar{\Phi}^2 \rangle - \langle \bar{\Phi} \rangle^2\,.
\end{eqnarray}

For vanishing quark chemical potential, $\chi_{ll}$ and $
\chi_{\bar{l}\bar{l}}$ coincide but are not equal to
$\chi_{l\bar{l}}$. In the low-temperature phase, where the $Z(3)$
symmetry is realized, the $Z(3)$ non-invariant components
$\chi_{ll}$ and $\chi_{\bar{l}\bar{l}}$ are strongly suppressed
\footnote{In purely gluonic theory, where the $Z(3)$ symmetry is
exact, $\chi_{ll}$ and $\chi_{\bar{l}\bar{l}}$ vanish identically
in the low temperature phase.}. In the $Z(3)$ symmetry broken phase
at high temperatures, all components of the Polyakov loop
susceptibility are large.

We also define the average susceptibility
\begin{equation}\label{average}
\bar{\chi}_{{l} {l}} =\frac{1}{4}(\chi_{ll}+\chi_{\bar{l}\bar{l}}+2\chi_{l \bar{l}})
%
\,.
\end{equation}
which corresponds to fluctuations of the real part of the Polyakov
loop.  This observable has been used in LGT calculations to
identify the position of deconfinement transition in the QCD
medium.

Due to the  presence of dynamical quarks in the PNJL model, the
$Z(3)$ symmetry is explicitly broken and the Polyakov loop is not a
genuine order parameter. Nevertheless, the fact that in LGT
calculations the expectation values of $\Phi$ and $\bar{\Phi}$ are
strongly suppressed in the low temperature phase~\cite{kac-zan},
suggests that the $Z(3)$ symmetry is a useful guiding principle for
constructing model Lagrangians (see also Figs.~\ref{fig:soln_Phi}
and~\ref{fig:soln-mu_Phi}).


\section{Susceptibilities and the phase transition in the PNJL
model} \label{sec:phase}

Enhanced fluctuations are characteristic for phase transitions.
Thus, the exploration of fluctuations is a promising tool for
probing the phase structure of a system. The phase boundaries can
be identified by the response of the fluctuations to changes in the
thermodynamic parameters. In this section we focus on the phase
structure of the PNJL model by studying the order parameter
susceptibilities, defined in the previous section.

\subsection{Susceptibilities at vanishing quark chemical potential}

\begin{figure}
\begin{center}
\includegraphics[width=8cm]{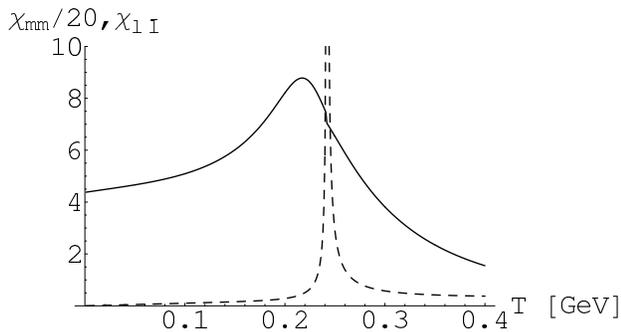}
\caption{\label{fig:sus_llb} The chiral $\chi_{mm}$ (dashed-line) and the Polyakov loop
$\chi_{l\bar{l}}$ (solid-line) susceptibilities in the chiral limit
as functions of temperature $T$ for $\mu = 0$. }
\end{center}
\end{figure}

In Fig.~\ref{fig:sus_llb} we show the chiral and Polyakov loop
susceptibilities $\chi_{mm}$ and $\chi_{l\bar{l}}$ computed at $\mu
= 0$ in the PNJL model in the chiral limit.
The chiral susceptibility exhibits a very narrow divergent peak at
the chiral critical temperature $T_{\rm ch}$, while the Polyakov
loop susceptibility shows a very different behavior: the peak is
much broader and the susceptibility remains finite for all
temperatures. The latter is due to the explicit breaking of the
$Z(3)$ symmetry by the presence of the quark fields in the PNJL
Lagrangian. Nevertheless, $\chi_{l\bar l}$ still exhibits a peak
structure that can be considered as a signal for the deconfinement
transition in this model.\footnote{As noted above, a similar
procedure is usually applied in LGT studies of QCD thermodynamics.}
The peak position of $\chi_{l\bar l}$ appears at $T\simeq 217$ MeV
below the chiral critical temperature $T_{\rm ch}\simeq 242$ MeV.
Thus, in this model the deconfinement phase transition occurs at a
lower temperature than that of the chiral phase transition. The
separation between the two peaks is roughly $25$ MeV in the
non-local model and about $8$ MeV in the local model.

Another interesting feature of $\chi_{l\bar l}$ is the interference
with the chiral susceptibility seen in Fig.~\ref{fig:sus_llb}. At
the chiral transition, $T=T_{\rm ch}$, there is a narrow structure
in $\chi_{l\bar l}$. We stress that this feature is not related
with the deconfinement transition, but only due to the coupling to
the chiral susceptibility. Thus, for the parameters used in the
model, the deconfinement transition, signaled by the broad bump in
$\chi_{l\bar l}$, sets in earlier than the chiral transition at
vanishing net quark density.


\subsection{Susceptibilities at a finite quark chemical potential}

At finite chemical potential, there is a shift of the chiral phase transition to lower
temperatures, as shown in Fig.~\ref{fig:phase_ch}. This is consistent with recent LGT
results at finite quark chemical potential~\cite{finite-mu}. A lowering of the
deconfinement temperature is also expected at non-zero net quark density. The position of
deconfinement and chiral transitions can be determined by exploring the order parameter
susceptibilities introduced in the previous section. With  increasing chemical potential
the temperature dependence of the Polyakov loop expectation value is flattening  and for
sufficiently large $\mu_q$  it shows almost no variation with $T$. Consequently, the
width of the  Polyakov loop susceptibility is increasing with  increasing  $\mu_q$.

\begin{figure}
\begin{center}
\includegraphics[width=8cm]{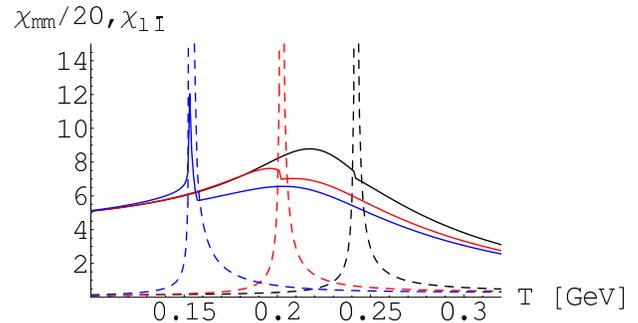}
\caption{\label{fig:sus_llb-mu} The chiral $\chi_{mm}$ (dashed) and  Polyakov loop
$\chi_{l\bar{l}}$ (solid) susceptibilities in the chiral limit as
functions of temperature $T$. The lines correspond to $\mu
= 0$ (right), $\mu = 200$ MeV (middle) and $\mu = 270$ MeV (left).
}
\end{center}
\end{figure}

\begin{figure}
\begin{center}
\includegraphics[width=8cm]{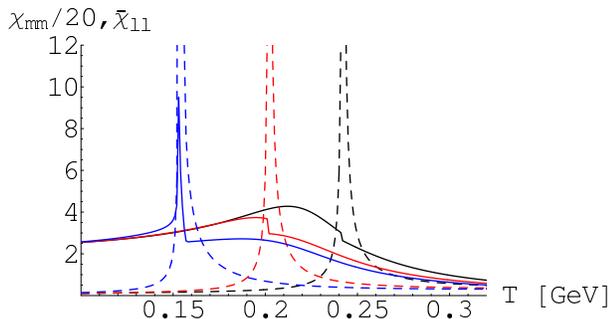}
\caption{\label{fig:sus_avr-mu} The chiral $\chi_{mm}$ (dashed) and Polyakov loop
$\bar{\chi}_{ll}$ (solid) susceptibilities in the chiral limit as
functions of temperature $T$. The lines correspond to $\mu
= 0$ (right), $\mu = 200$ MeV (middle) and $\mu = 270$ MeV (left).
}
\end{center}
\end{figure}

In Figs.~\ref{fig:sus_llb-mu} and \ref{fig:sus_avr-mu} we
illustrate the temperature dependence of the susceptibilities for
several values of the chemical potential. With increasing $\mu$ the
peak position of the chiral and Polyakov loop susceptibilities are
clearly shifted towards lower $T$ and approach each other as seen
in Fig.~\ref{fig:sus_llb-mu}.
At $\mu_0\simeq 185$ MeV  the two peaks coincide, which indicates
that the chiral and deconfinement transitions appear at the same
temperature. For $\mu>\mu_0$ the Polyakov loop susceptibility
$\chi_{l\bar{l}}$ has a sharp peak at $T_{\rm ch}$ and a broad bump
above $T_{\rm ch}$. A similar behavior was also found in
Ref.~\cite{Fukushima}. The peak at $T_{\rm ch}$ in
$\chi_{l\bar{l}}$ is clearly due to an interference with the chiral
phase transition, while the bump corresponds to the pseudo-critical
point of the deconfinement transition. This structure is also seen
in $\chi_{l\bar{l}}$ at the tricritical point, located at $(T_c=157
,\mu_c=266 )$ MeV.

In the previous section we also introduced the average
susceptibility $\bar\chi_{ll}$, which corresponds to fluctuations
of the real part of the Polyakov loop. In Fig.~\ref{fig:sus_avr-mu}
the temperature variation of $\bar\chi_{ll}$ for different values
of the quark chemical potential is shown. It is clear from this
figure that a behavior of $\bar\chi_{ll}$  is very similar to that
of $\chi_{l\bar l}$. However, the peak positions of the two
susceptibilities differ slightly. Thus, in the Polyakov loop sector
the determination of the transition temperature by using the
susceptibilities is not unique. Nevertheless, the differences are
small, so in the following we use $\chi_{l\bar l}$ to identify the
deconfinement transition temperature in PNJL model.

\begin{figure}
\begin{center}
\includegraphics[width=8cm]{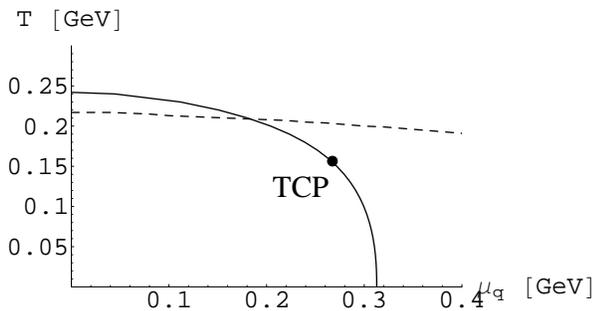}
\caption{\label{fig:phase1} The phase diagram of the PNJL model in the chiral limit. The
solid (dashed) line denotes the chiral (deconfinement) phase
transition respectively. The TCP (bold-point) is located at
$(T_c=157 ,\mu_c=266)$ MeV. The parameter set (a) in
Table~\ref{table:Tc_adj} was used in the calculation. }
\end{center}
\end{figure}

\begin{figure}
\begin{center}
\includegraphics[width=8cm]{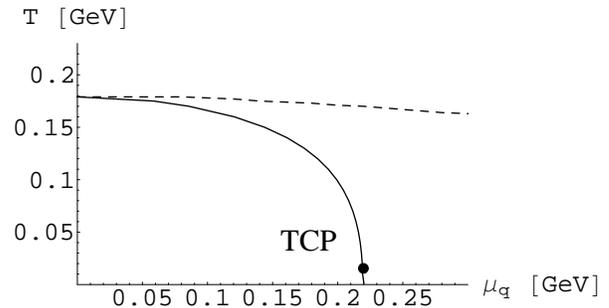}
\caption{\label{fig:phase2} The phase diagram of the PNJL model in the chiral limit. The
solid (dashed) line denotes the chiral (deconfinement) phase
transition respectively. The TCP (bold-point) is located at
$(T_c=15,\mu_c=218)$ MeV. The results correspond to parameter set
(b) in Table~\ref{table:Tc_adj}. }
\end{center}
\end{figure}

The peak positions of the $\chi_{mm}$ and $\chi_{l\bar{l}}$
susceptibilities are used to determine the phase boundaries in the
$(T,\mu)$--plane. We stress that the phase boundaries are strongly
dependent on the model parameters. In Fig.~\ref{fig:phase1} we show
the phase diagram for the PNJL model obtained with the parameters
from Table~\ref{table:njl}. Clearly the boundary line related with
the chiral symmetry restoration, determined by the peak in the
chiral susceptibility, coincides with that in
Fig.~\ref{fig:phase_ch}, which was computed from the properties of
the dynamical quark mass. Furthermore, the phase boundary
corresponding to the deconfinement transition is identified with
the position of the broad maximum observed in $\chi_{l\bar{l}}$.
With these parameters the boundary lines of deconfinement and
chiral symmetry restoration do not coincide. As anticipated in the
discussion of Fig.~\ref{fig:sus_llb-mu}, there is only one common
point in the phase diagram where the two transitions appear
simultaneously.

Recent LGT results both at vanishing and at finite quark chemical
potential show that deconfinement and chiral symmetry restoration
appears in QCD along the similar critical line \cite{lattice:sus}.
In general it is possible to choose the PNJL model parameters such
that the critical temperatures of chiral and deconfinement
transition coincide at $\mu=0$. The resulting phase diagram is
shown in Fig.~\ref{fig:phase2}; the parameters used in the
calculations are summarized in Table~\ref{table:Tc_adj}, set (b).
We note that this choice of the parameters is not unique. A shift
in the position of the critical temperature at $\mu=0$ can also be
obtained by changing the $T_0$-parameter in the effective gluon
potential. Decreasing $T_0$ from 270 MeV to 130 MeV results in
$T_{\rm ch}=192$ MeV, consistent with recent LGT calculations
\cite{karsch}.

From Figs.~\ref{fig:phase1} and~\ref{fig:phase2} it is clear that
in our model there is only a rather narrow region of finite $\mu$
where the deconfinement and chiral transition lines coincide. The
slope of $T_{\rm dec}$ as a function of $\mu$ is almost flat,
indicating that at low temperature the chiral phase transition
should appear much earlier than deconfinement. So far there is no
guidance available from first principle LGT calculations concerning
the relation between deconfinement and chiral symmetry restoration
at large values of the chemical potential. However, there are
general arguments, that the deconfinement transition should precede
restoration of chiral symmetry (see e.g. \cite{Shuryak, Pisarski}).
In view of this, it seems unlikely that at $T\simeq 0$ the chiral
symmetry sets in at the lower baryon density than deconfinement. In
the PNJL model, the effective gluon potential parameters were fixed
by fitting quenched LGT calculations. Consequently, the parameters
are taken as independent on $\mu$. However, it is conceivable  that
the effect of dynamical quarks can modify the coefficients of this
potential, thus resulting in $\mu$--dependence of the parameters.
Consequently, the slope of $T_{\rm dec}$ as a function of $\mu$
could be steeper~\footnote{Such a modification was explored in
Ref.~\cite{mu-dep} where explicit $\mu$- and $N_f$- dependence of
$T_0$ is extracted from the running coupling constant $\alpha_s$,
using the argument based on the renormalization group.}.
Consequently, the effective Polyakov loop potential
(\ref{eff_potential}) should, with $\mu$-independent coefficients,
be employed only for $\mu/T<1$.

\begin{figure}
\begin{center}
\includegraphics[width=8cm]{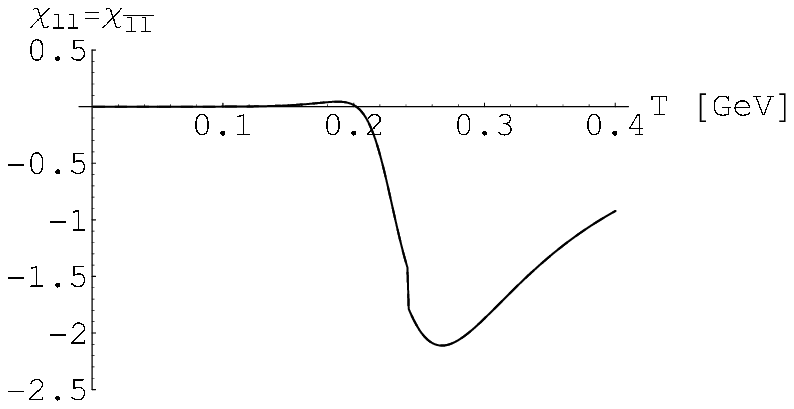}
\caption{\label{fig:sus_ll} The diagonal $\chi_{ll}=\chi_{\bar{l}\bar{l}}$ susceptibility
in the chiral limit as a function of temperature $T$ for $\mu = 0$. }
\end{center}
\end{figure}
\begin{figure}
\begin{center}
\includegraphics[width=8cm]{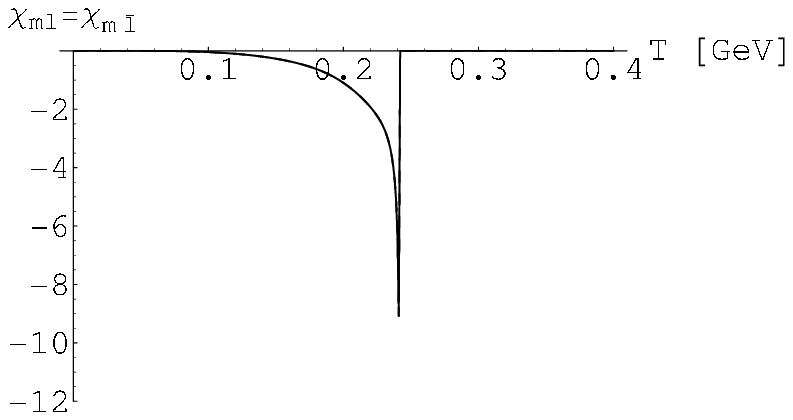}
\caption{\label{fig:sus_ml} The off-diagonal $\chi_{ml} = \chi_{m\bar{l}}$ susceptibility
in the chiral limit as a function of temperature $T$ for $\mu = 0$. }
\end{center}
\end{figure}
In Figs.~\ref{fig:sus_ll} and~\ref{fig:sus_ml} we show further
diagonal and off-diagonal components of the susceptibility
$\hat{\chi}$ (\ref{chi-matrix}) at $\mu = 0$. In this case the
diagonal components $\chi_{ll}$ and $\chi_{\bar{l}\bar{l}}$
coincide. These are the $Z(3)$ non-invariant susceptibilities for
$\Phi$ and $\bar{\Phi}$. These components are suppressed in the
low-temperature phase, where the $Z(3)$ symmetry is (approximately)
restored. Around $T\simeq T_{\rm dec}$ the
$\chi_{ll,\bar{l}\bar{l}}$ show a rapid drop, associated with the
crossover transition, where the expectation values $\Phi$ and
$\bar{\Phi}$ grow rapidly. In the high-temperature phase
$\chi_{ll,\bar{l}\bar{l}}$ are not necessarily suppressed since the
$Z(3)$ symmetry is explicitly broken there. The off-diagonal
susceptibilities $\chi_{ml} =
\chi_{m\bar{l}}$ are also suppressed in the low-temperature phase and show a clear peak
at $T_{\rm ch}$ due to an interference with the chiral phase
transition. These susceptibilities are non-invariant under both the
$Z(3)$ and chiral symmetries. Hence, they are suppressed well below
the transition, where the $Z(3)$ symmetry is restored and vanish
above $T_{\rm ch}$, in the chirally symmetric phase.

\begin{figure*}
\begin{center}
\includegraphics[width=8cm]{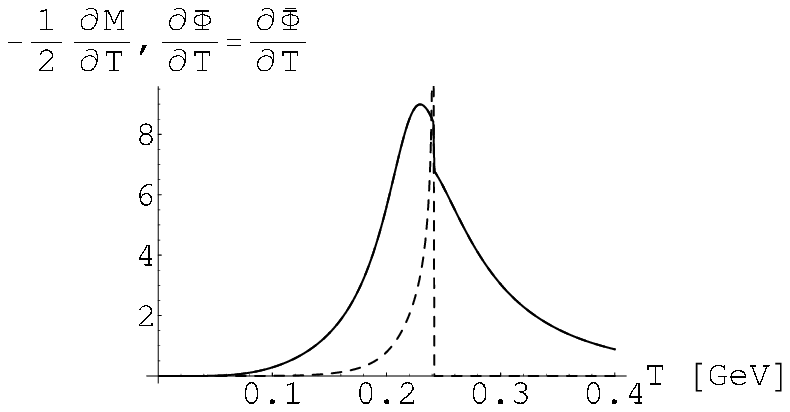}
\hspace*{0.5cm}
\includegraphics[width=8cm]{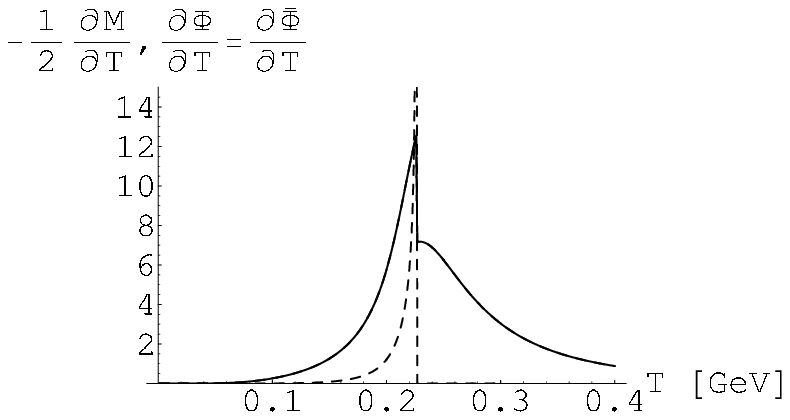}
\\
(a) $T_0 = 270$ MeV, non-local
\hspace*{4.5cm}
(b) $T_0 = 270$ MeV, local
\\
\includegraphics[width=8cm]{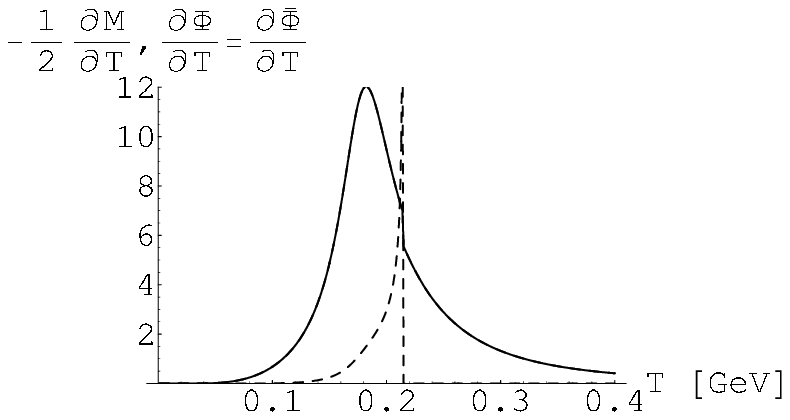}
\hspace*{0.5cm}
\includegraphics[width=8cm]{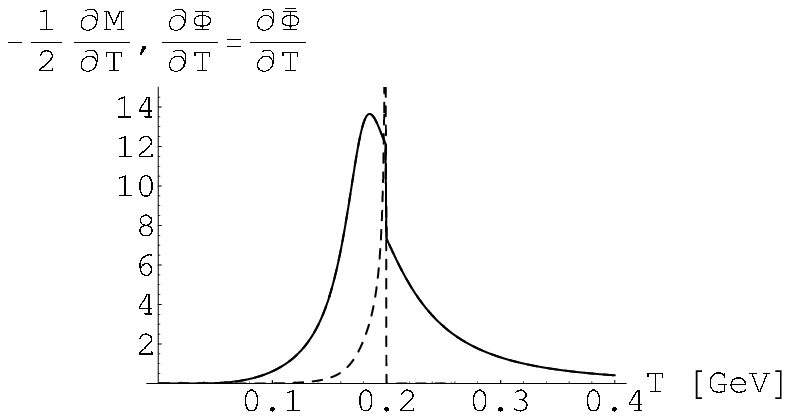}
\\
(c) $T_0 = 210$ MeV, non-local \hspace*{4.5cm} (d) $T_0 = 210$ MeV, local
\caption{\label{fig:der-mu0} Derivatives of the dynamical quark mass $\partial M/\partial
T$ (dashed) and the Polyakov loop $\partial \Phi/\partial T$
(solid) in the chiral limit as functions of temperature $T$ for
$\mu= 0$. }
\end{center}
\end{figure*}

An alternative way to determine the deconfinement and chiral
transition temperatures, by locating the maximum of the temperature
derivative of the corresponding effective condensate, has been
discussed in the literature (see e.g. ref.~\cite{RRW:phase}). The
temperature derivatives of the condensates can be expressed as
combinations of the susceptibilities with some $T$- and
$\mu$-dependent coefficients, as shown in Eq.~(\ref{T-der}).
Consequently, in general the deconfinement pseudo critical
temperature obtained with this method does not agree with that
obtained from the peak position of the corresponding
susceptibilities. Only if the phase transitions are relatively
sharp and the susceptibilities show narrow structures can one
expect that the transition temperatures determined using the
different prescriptions coincide. In Fig.~\ref{fig:der-mu0} we show
the derivatives of order the parameters at $\mu = 0$  for different
values of $T_0$.

In ref.~\cite{PNJL} the peak positions in the derivatives of the
Polyakov loop and of the chiral order parameter coincide. Indeed,
in the local version of the PNJL model with $T_0
= 270$ MeV we reproduce this result, as illustrated in Fig.~\ref{fig:der-mu0}.
However, for a slightly different value of $T_0$, the positions of
the two peaks split. The parameters in the Polyakov loop sector
were fixed from the lattice data in the heavy quark mass limit.
This corresponds to a transition temperature $T_0
= 270$ MeV in the pure gauge theory. In the presence of dynamical
quarks the transition temperature drops, thus it is not excluded
that the value of  $T_0$ may depend on $N_f$, as noted in the
introduction. With $T_0 = 210$ MeV, the local PNJL model yields a
splitting  of the peaks in the derivatives of the Polyakov loop and
the chiral condensate. Thus, the locations of the pseudo-critical
points are strongly dependent on the model parameters. In the
non-local PNJL model considered in this work, there is no
coincidence between the peak positions and the peak  of
$\frac{\partial
\Phi}{\partial T}$ is located at a lower temperature than that of
$\frac{\partial M}{\partial T}$ for all values of $T_0 \lapp 300$
MeV, as illustrated in Fig.~\ref{fig:der-mu0}a and c. We note that
in the non-local PNJL model the chiral transition temperatures
obtained, on the one hand, from the derivative of the order
parameter and, on the other hand, from the corresponding
susceptibility are almost identical. Thus, at least in the chiral
limit, the two methods for identifying the chiral transition, are
equivalent. However, for the Polyakov loop, the difference is $\sim
10$ MeV at $\mu = 0$.


\subsection{ Effective potential constraints }

As shown in Fig.~\ref{fig:sus_ll}, $\chi_{{l} {l}}$ is negative in
a broad temperature range above $T_{\rm ch}$. This is in
disagreement with recent lattice results, where $\chi_{{l} {l}}$ is
always positive in the presence of dynamical quarks
\cite{lattice:sus}. Furthermore, the relation of the Polyakov loop
susceptibility with the free energy of static quarks is also
incompatible with negative values of $\chi_{{l} {l}}$. A possible
origin of this behavior could be the approximation to the effective
Polyakov loop potential used in the Eq. (\ref{eff_potential}).

The $\chi_{{l} {l}}$ susceptibility can be related to the screening
masses $m_r$ and $m_i$ of the real and imaginary part of the
Polyakov loop correlation functions \cite{poly:mass}:
\begin{equation}
\chi_{ll} = \frac{1}{m_r^2} - \frac{1}{m_i^2}\,.\label{mm}
\end{equation}
In Ref.~\cite{poly:mass}  the   ratio
\begin{equation}
\frac{m_i}{m_r} \simeq 3\,\label{mm1}
\end{equation}
was found in the vicinity to the critical temperature. This result
is qualitatively consistent with the perturbative
calculation~\cite{mass:pert}. From  the  Eqs. (\ref{mm}) and
(\ref{mm1}) it is clear that the resulting $\chi_{ll}$ should  be
positive at $T=T_c$.

In Ref.~\cite{poly:mass}  the screening masses $m_r$ and $m_i$ were
calculated with an effective potential that has the same polynomial
form as used in the Eq. (\ref{eff_potential}) but with  different
values of the parameters. With the values given in table
\ref{table:pol} effective Polyakov loop potential
(\ref{eff_potential}) yields
\begin{equation}
m_i < m_r\,.
\end{equation}
which implies a negative value of $\chi_{ll}$ at $T_c$.

From this discussion  it is clear that the behavior   of
$\chi_{ll}$ depends crucially on the parameters used in the
Polyakov loop potential. Thus, the constraints from $Z(3)$ symmetry
and from the lattice results for the equation of state, are not
sufficient to warrant physically acceptable susceptibilities.

Recently an improved effective potential with temperature-
dependent coefficients has been suggested ~\cite{RRW:phase}
\begin{eqnarray}
&&
\frac{{\cal U}(\Phi,\bar{\Phi};T)}{T^4}
= - \frac{a(T)}{2} \bar{\Phi}\Phi
\nonumber\\
&&
{}+ b(T) \ln \left[ 1 - 6\bar{\Phi}\Phi
{}+ 4\left( \Phi^3 + \bar{\Phi}^3 \right) - 3\left( \bar{\Phi}\Phi \right)^2 \right]\,,
\label{eff_imp}
\end{eqnarray}
where
\begin{equation}
a(T) = a_0 + a_1 \left( \frac{T_0}{T} \right) + a_2 \left( \frac{T_0}{T} \right)^2\,,
\quad
b(T) = b_3  \left( \frac{T_0}{T} \right)^3\,,
\end{equation}
The polynomial in $\Phi$ and $\bar{\Phi}$,  used in
(\ref{eff_potential}), is replaced by a logarithmic term, which
accounts for the Haar measure in the group
integral~\cite{Fukushima}. The parameters in (\ref{eff_imp}) were
fixed to reproduce the lattice results for pure gauge QCD
thermodynamics and for the behavior of the Polyakov loop. These
parameters are summarized in the Table~\ref{table:imp}.

\begin{figure}
\begin{center}
\includegraphics[width=8cm]{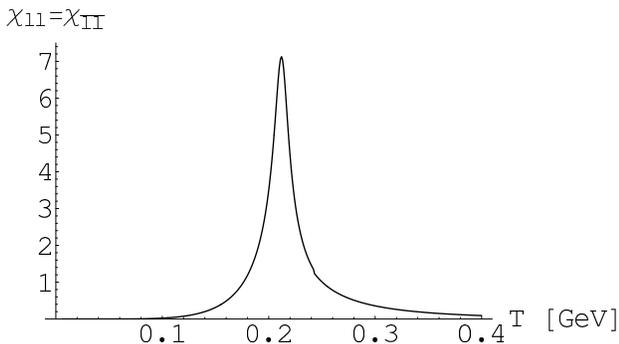}
\caption{\label{fig:susll_imp} The diagonal $\chi_{ll}=\chi_{\bar{l}\bar{l}}$ susceptibility
in the chiral limit as a function of temperature $T$ for $\mu = 0$.
The effective Polyakov loop potential (\ref{eff_imp}) was used.}
\end{center}
\end{figure}

\begin{figure}
\begin{center}
\includegraphics[width=8cm]{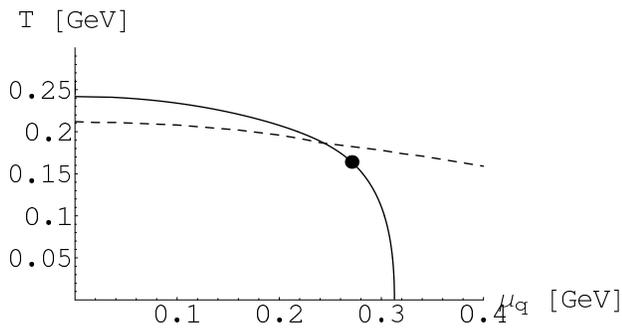}
\caption{\label{fig:phase_imp} The phase diagram of the PNJL model in the chiral limit. The
solid (dashed) line denotes the chiral (deconfinement) phase transition respectively. The
TCP (bold-point) is located at $(T_c=164 ,\mu_c=271)$ MeV. The parameter set (a) given
in Table~\ref{table:Tc_adj} was used in the calculations.
}
\end{center}
\end{figure}

In Fig.~\ref{fig:susll_imp} we show the  $\chi_{ll}$ susceptibility calculated with the
potential of Eq. (\ref{eff_imp}).  It is clear from this figure that the improved
potential yields positive values for the Polyakov loop susceptibilities . In addition the
peak positions of the $\chi_{ll}$ and $\chi_{l\bar{l}}$ susceptibilities almost coincide
if the effective Polyakov loop potential is parameterized as in Eq. (\ref{eff_imp}). The
phase diagram calculated with an improved potential, shown in Fig.~\ref{fig:phase_imp},
is similar to that obtained with the previous choice of the Polyakov loop interactions,
Fig.~\ref{fig:phase1}. This is also the case for the  susceptibilities $\chi_{mm}$ and
$\chi_{l\bar l}$ which have  the same   qualitative structure  for both effective
potentials.

\setcounter{equation}{0}
\section{Summary and Conclusions}
\label{sec:sum}

We have explored the thermodynamical properties and the critical
behavior of a system that exhibits  an invariance under $Z(3)$ and
chiral transformations. As a model we have use an extension of the
NJL model for quarks with three colors and two flavors, which are
coupled to an effective gluon field described by the Polyakov loop.
In this model (the PNJL model) a non-local interaction instead of
the point-like four--fermion couplings is employed. The PNJL model
exhibits two essential  features of QCD: a spontaneous chiral
symmetry breaking and a ``confinement''like property. Furthermore,
due to the symmetries of the Lagrangian, the PNJL model belongs to
the same universality class as that expected for QCD. Thus, such a
model can be considered as a testing ground for the critical
phenomena related to the breaking of the global $Z(3)$ and chiral
symmetries.

Within the PNJL model, we discussed the phase diagram and the order
of the phase transition, using mean field dynamics for different
values of the parameters. The properties of thermodynamic
quantities related with the quark degrees of freedom, like the
quark number density and susceptibility, were analyzed in the
vicinity of the chiral and deconfinement transitions.

We introduced susceptibilities related with the three different
order parameters in this model, and analyzed their properties and
their behavior near the phase transitions. We have shown that
there are as many as nine susceptibilities that can be used to
identify the phase structure of the model. In particular, for the
quark-antiquark and chiral density-density correlations we have
discussed the interplay between the restoration of chiral symmetry
and deconfinement. We argued that in the actual formulation of the
PNJL model a coincidence of the deconfinement and chiral symmetry
restoration is accidental.

We found that, within the mean field approximation and with the
present form of an effective gluon potential the correlations of
the Polyakov loops in the quark--quark channel show an unphysical
behavior, being negative in a broad parameter range. This behavior
was traced back to the parameterization  of the Polyakov loop
potential. We argue that the $Z(N)$-invariance of this potential
and the fit to lattice thermodynamics in the pure gluon sector is
not sufficient to provide correct description of the Polyakov loop
fluctuations in the presence of quarks in a medium. We note,
however, that the improved potential of ref.~\cite{RRW:phase}
yields a positive, i.e. physical, $\chi_{ll}$, in qualitative
agreement with the LGT results.


\section*{Acknowledgments}
We acknowledge stimulating discussions with D. Blaschke, A. Dumitru, S. Ejiri, F. Karsch,
H.~Gies, J. Pawlowski, H.~J.~Pirner, C. Ratti,
B.~J.~Schaefer, D. Voskresensky, J. Wambach and W. Weise. K.R. acknowledges fruitful
discussion with R.D. Pisarski.
 The work of B.F. and C.S. was supported in part
by the Virtual Institute of the Helmholtz Association under the grant No. VH-VI-041. K.R.
acknowledges partial support of the Gesellschaft f\"ur Schwerionenforschung (GSI) and KBN
under grant 2P03 (06925).


\appendix

\setcounter{section}{0}
\renewcommand{\thesection}{\Alph{section}}
\setcounter{equation}{0}
\renewcommand{\theequation}{\Alph{section}.\arabic{equation}}

\setcounter{equation}{0}

\setcounter{equation}{0}
\section{Functional integral and partition function of the model}
\label{app:basis}

The Polyakov loop $L$  is a complex $3\times 3$ matrix and can be
diagonalized as in Eq.~(\ref{Lmatrix}). Thus, the thermodynamic
potential  (\ref{omega2}) is a functional of complex variables.
However, as a physical observable the imaginary part of $\Omega$
must vanish. In this Appendix we show that this indeed happens. We
follow the method that has been used in the context of a strong
coupling QCD and the matrix model ~\cite{miller,matrix-model}.

We start with the partition function,
\begin{equation}
Z = \int {\mathcal D}\psi {\mathcal D}\bar{\psi}
{\mathcal D}L(\varphi,\varphi^\prime)
e^{S[\psi,\bar{\psi},L]}\,,
\end{equation}
with the action $S$ being  divided into three pieces:
\begin{eqnarray}
S[\psi,\bar{\psi},L]
&=&
\int_0^\beta d\tau \int d^3x {\mathcal L}
[\psi,\bar{\psi},L]
\nonumber\\
&=&
S_g [L] + S_q [\psi,\bar{\psi},L]
{}+ S_{\rm int}[\psi,\bar{\psi}]\,.
\end{eqnarray}
The Polyakov loop effective potential ${\mathcal U}$ is included in $S_g$. It is quite
transparent that   the pure gluon part is real. Thus, we  focus only on the quark part
$S_q$. Performing the functional integral over fermion fields one gets,
\begin{equation}
Z_q = \int {\mathcal D}L(\varphi,\varphi^\prime)\,
{\mbox{Det}}D(\varphi,\varphi^\prime)\,,
\end{equation}
where $D$ denotes the Dirac operator
\begin{equation}
D(\varphi,\varphi^\prime)
= \Slash{p} - M - \gamma_0 (\mu + iA_4)\,.
\end{equation}
\begin{widetext}
Summing up  the  Matsubara frequencies, $p_0 = i\omega_n = i(2n + 1)\pi T$ the partition
function is obtained as
\begin{eqnarray}
\ln Z_q
&=&
\int {\mathcal D}L(\varphi,\varphi^\prime)
\sum_n \int \frac{d^3 p}{(2\pi)^3}
\mbox{Tr}\ln D(\varphi,\varphi^\prime)
\nonumber\\
&=&
2 N_f \int {\mathcal D}L(\varphi,\varphi^\prime)
\int \frac{d^3 p}{(2\pi)^3}
\Bigl[
\mbox{Tr}_c \ln \left( 1 + L(\varphi,\varphi^\prime)
e^{-\beta E^{(+)}} \right)
{}+ \mbox{Tr}_c \ln \left( 1 + L^\dagger(\varphi,\varphi^\prime)
e^{-\beta E^{(-)}} \right)
\Bigr]\,,
\nonumber\\
\end{eqnarray}
taking the  trace over color, flavor and Dirac variables in the above equation the
imaginary part  $I$ of $\ln Z_q$ is proportional to
\begin{eqnarray}
I(M,\Phi,\bar{\Phi};T,\mu)
&=&
\int {\mathcal D}L(\varphi,\varphi^\prime)
\int \frac{d^3 p}{(2\pi)^3}
\Biggl\{\tan^{-1}\left[ \frac{3\mbox{Im}\Phi
\left( 1 - e^{-E^{(+)}/T} \right)e^{-E^{(+)}/T}}
{1 + 3\mbox{Re}\Phi \left( 1 + e^{-E^{(+)}/T} \right)
e^{-E^{(+)}/T} + e^{-3E^{(+)}/T}} \right]
\nonumber\\
&&\qquad{}+
\tan^{-1}\left[ \frac{- 3\mbox{Im}\Phi
\left( 1 - e^{-E^{(-)}/T} \right)e^{-E^{(-)}/T}}
{1 + 3\mbox{Re}\Phi \left( 1 + e^{-E^{(-)}/T} \right)
e^{-E^{(-)}/T} + e^{-3E^{(-)}/T}} \right]
\Biggr\}\,,
\label{im}
\end{eqnarray}
where $\varphi$ and $\varphi^\prime$ dependence were suppressed and $\mbox{Re}\Phi =
\mbox{Re}\bar{\Phi}$ and $\mbox{Im}\Phi = - \mbox{Im}\bar{\Phi}$ were used. Now let us
replace the variables $\varphi, \varphi^\prime$ with $\varphi \to - \varphi$ and
$\varphi^\prime \to - \varphi^\prime$. The SU(3) Haar measure ${\mathcal
D}L(\varphi,\varphi^\prime)$ is unchanged under these replacements  while $\mbox{Im}\Phi$
changes its sign. Therefore, the first and second terms are separately odd under the change of
group variables and vanish after the integration.

The thermal expectation values of complex $\Phi$ and $\bar{\Phi}$
are evaluated as
\begin{eqnarray}
\langle \Phi \rangle
&=&
\frac{1}{Z}\int {\mathcal D}L(\varphi,\varphi^\prime)
e^{S_g + S_{\rm int}}\left( \mbox{Re}z_q \cdot \mbox{Re}\Phi
{}- \mbox{Im}z_q \cdot \mbox{Im}\Phi \right)\,,
\nonumber\\
\langle \bar{\Phi} \rangle
&=&
\frac{1}{Z}\int {\mathcal D}L(\varphi,\varphi^\prime)
e^{S_g + S_{\rm int}}\left( \mbox{Re}z_q \cdot \mbox{Re}\Phi
{}+ \mbox{Im}z_q \cdot \mbox{Im}\Phi \right)\,,
\end{eqnarray}
with
\begin{equation}
Z = \int {\mathcal D}L(\varphi,\varphi^\prime)
e^{S_g + S_{\rm int}}\left( \mbox{Re}z_q + i\mbox{Im}z_q \right)\,.
\end{equation}
It is clear  from Eq.~(\ref{im}) that the imaginary part of the potential vanishes at
$\mu = 0$. Thus,  the difference between $\langle \Phi \rangle$ and $\langle \bar{\Phi}
\rangle$ comes only from the non-vanishing $\mbox{Im}z_q$ at finite $\mu$. This can be
also seen in the matrix model for color SU(3)
symmetry~\cite{matrix-model}.


\setcounter{equation}{0}
\section{Derivatives of effective condensates}
\label{app:der}

 In this appendix we summarize  the derivatives of effective condensates
with respect to $\mu$ and $T$. Taking the $\mu$-derivatives in the coupled gap equations
(\ref{gapeq:M})-(\ref{gapeq:Phi-bar}) one gets
\begin{eqnarray}
\frac{\partial M}{\partial\mu}
&=&
\frac{T}{\Lambda}\left( A_M^{(\mu)}\chi_{mm}
{}+ \frac{T}{\Lambda}A_\Phi^{(\mu)}\chi_{ml}
{}+ \frac{T}{\Lambda}A_{\bar{\Phi}}^{(\mu)}\chi_{m\bar{l}}
\right)\,,
\nonumber\\
\frac{\partial\Phi}{\partial\mu}
&=&
\frac{T}{\Lambda^2}\left( A_M^{(\mu)}\chi_{ml}
{}+ \frac{T}{\Lambda}A_\Phi^{(\mu)}\chi_{ll}
{}+ \frac{T}{\Lambda}A_{\bar{\Phi}}^{(\mu)}\chi_{l\bar{l}}
\right)\,,
\nonumber\\
\frac{\partial\bar{\Phi}}{\partial\mu}
&=&
\frac{T}{\Lambda^2}\left( A_M^{(\mu)}\chi_{m\bar{l}}
{}+ \frac{T}{\Lambda}A_\Phi^{(\mu)}\chi_{l\bar{l}}
{}+ \frac{T}{\Lambda}A_{\bar{\Phi}}^{(\mu)}\chi_{\bar{l}\bar{l}}
\right)\,,
\label{mu-der}
\end{eqnarray}
where $\chi_{ij}$ are defined in Section~\ref{sec:def_sus} and the functions $A^{(\mu)}$
are introduced as
\begin{eqnarray}
A_M^{(\mu)}
&=&
- \frac{6 N_f}{T^3} \int\frac{d^3 p}{(2\pi)^3}
\frac{M_p f^2(p)}{E_p}\Biggl[
\frac{e^{-E^{(+)}/T}}{(g^{(+)})^2}\Bigl\{
\Phi + 4\bar{\Phi}e^{-E^{(+)}/T}
{}+ 3 \left( 1 + \bar{\Phi}\Phi \right) e^{-2E^{(+)}/T}
\nonumber\\
&&
{}+ 4\Phi e^{-3E^{(+)}/T}
{}+ \bar{\Phi}e^{-4E^{(+)}/T}
\Bigr\}
{}- \left( \bar{\Phi},\Phi;-\mu \right)
\Biggr]\,,
\nonumber\\
A_\Phi^{(\mu)}
&=&
\frac{6 N_f}{T^3} \int\frac{d^3 p}{(2\pi)^3}
\Biggl[
\frac{e^{-E^{(+)}/T}}{(g^{(+)})^2}\Bigl\{
1 - 3\bar{\Phi}e^{-2E^{(+)}/T} - 2 e^{-3E^{(+)}/T}
\Bigr\}
\nonumber\\
&&
{}-
\frac{e^{-2E^{(-)}/T}}{(g^{(-)})^2}\Bigl\{
2 + 3\bar{\Phi}e^{-E^{(-)}/T} - e^{-3E^{(-)}/T}
\Bigr\}
\Biggr]\,,
\nonumber\\
A_{\bar{\Phi}}^{(\mu)}
&=&
A_\Phi^{(\mu)}(\bar{\Phi},\Phi;-\mu)\,.
\end{eqnarray}

The required temperature derivatives of order parameters are  directly obtained from the
gap equations  as
\begin{eqnarray}
\frac{\partial M}{\partial T}
&=&
\frac{T}{\Lambda}\left( A_M^{(T)}\chi_{mm}
{}+ \frac{T}{\Lambda}A_\Phi^{(T)}\chi_{ml}
{}+ \frac{T}{\Lambda}A_{\bar{\Phi}}^{(T)}\chi_{m\bar{l}}
\right)\,,
\nonumber\\
\frac{\partial\Phi}{\partial T}
&=&
\frac{T}{\Lambda^2}\left( A_M^{(T)}\chi_{ml}
{}+ \frac{T}{\Lambda}A_\Phi^{(T)}\chi_{ll}
{}+ \frac{T}{\Lambda}A_{\bar{\Phi}}^{(T)}\chi_{l\bar{l}}
\right)\,,
\nonumber\\
\frac{\partial\bar{\Phi}}{\partial T}
&=&
\frac{T}{\Lambda^2}\left( A_M^{(T)}\chi_{m\bar{l}}
{}+ \frac{T}{\Lambda}A_\Phi^{(T)}\chi_{l\bar{l}}
{}+ \frac{T}{\Lambda}A_{\bar{\Phi}}^{(T)}\chi_{\bar{l}\bar{l}}
\right)\,,
\label{T-der}
\end{eqnarray}
where the functions $A^{(T)}$ are defined as
\begin{eqnarray}
A_M^{(T)}
&=&
- \frac{6 N_f}{T^4} \int\frac{d^3 p}{(2\pi)^3}
\frac{M_p f^2(p)}{E_p}\Biggl[
\frac{e^{-E^{(+)}/T}}{(g^{(+)})^2}E^{(+)}
\Bigl\{
\Phi + 4\bar{\Phi}e^{-E^{(+)}/T}
{}+ 3 \left( 1 + \bar{\Phi}\Phi \right) e^{-2E^{(+)}/T}
\nonumber\\
&&
{}+ 4\Phi e^{-3E^{(+)}/T}
{}+ \bar{\Phi}e^{-4E^{(+)}/T}
\Bigr\}
{}+ \left( \bar{\Phi},\Phi;-\mu \right)
\Biggr]\,,
\nonumber\\
A_\Phi^{(T)}
&=&
\frac{T}{2}\frac{\partial b_2}{\partial T}\bar{\Phi}
{}+ \frac{6 N_f}{T^4} \int\frac{d^3 p}{(2\pi)^3}
\Biggl[
\frac{e^{-E^{(+)}/T}}{(g^{(+)})^2}E^{(+)}
\Bigl\{
1 - 3\bar{\Phi}e^{-2E^{(+)}/T} - 2 e^{-3E^{(+)}/T}
\Bigr\}
\nonumber\\
&&
{}-
\frac{e^{-2E^{(-)}/T}}{(g^{(-)})^2}E^{(-)}
\Bigl\{
2 + 3\bar{\Phi}e^{-E^{(-)}/T} - e^{-3E^{(-)}/T}
\Bigr\}
\Biggr]
{}- \frac{18 N_f}{T^3}\int\frac{d^3 p}{(2\pi)^3}
\Biggl[
\frac{e^{-E^{(+)}/T}}{g^{(+)}} + \frac{e^{-2E^{(-)}/T}}{g^{(-)}}
\Bigg]\,,
\nonumber\\
A_{\bar{\Phi}}^{(T)}
&=&
A_\Phi^{(T)}(\bar{\Phi},\Phi;-\mu)\,.
\end{eqnarray}


\section{Parameters used in the PNJL model thermodynamics}
\label{tables}

The compilation of parameters used in the model calculations is given in the following
tables:
\begin{table}[ht]
\begin{center}
\begin{tabular*}{14cm}{@{\extracolsep{\fill}}c|ccc}
\hline
input & $f_\pi = 92.4$ MeV & $m_\pi = 135$ MeV & $M = 335$ MeV\\
\hline\hline non-local NJL model & $\Lambda = 684.2$ MeV &
$G_S \Lambda^2 = 4.66$ & $m_0 = 4.46$ MeV\\
\hline local NJL model & $\Lambda = 625.1$ MeV &
$G_S \Lambda^2 = 4.38$ & $m_0 = 5.31$ MeV\\
\hline
\end{tabular*}
\end{center}
\caption{
 Set of parameters for the NJL sector~\cite{nl:parameters}.
 The parameters of non-local NJL model were fixed
 for $\alpha = 10$.
} \label{table:njl}
\end{table}

\begin{table}[ht]
\begin{center}
\begin{tabular*}{9cm}{@{\extracolsep{\fill}}cccccc}
\hline
$a_0$  & $a_1$   & $a_2$   & $a_3$   & $b_3$  & $b_4$\\
\hline
$6.75$ & $-1.95$ & $2.625$ & $-7.44$ & $0.75$ & $7.5$ \\
\hline
\end{tabular*}
\end{center}
\caption{ Set of parameters for the Polyakov-loop effective potential~\cite{PNJL}. }
\label{table:pol}
\end{table}

\begin{table}[ht]
\begin{center}
\begin{tabular*}{8cm}{@{\extracolsep{\fill}}lc||lc}
\hline
set (a) & {\quad} & set (b) & {\quad}\\
\hline
$\Lambda = 684.2$ MeV & {} & $\Lambda = 684.2$ MeV & {}\\
$G_S \Lambda^2 = 4.66$ & {} & $G_S \Lambda^2 = 4.05$ & {}\\
$T_0 = 270$ MeV & {} & $T_0 = 225$ MeV & {}\\
\hline
$T_{\rm ch}(\mu=0) = 242$ MeV & {} & $T_{\rm ch}(\mu=0) = 180$ MeV & {}\\
$T_{\rm dec}(\mu=0) = 217$ MeV & {} & $T_{\rm dec}(\mu=0) = 180$ MeV & {}\\
\hline
\end{tabular*}
\end{center}
\caption{
 Set of parameters in the chiral limit used in this work and the resultant
 phase transition temperatures.
 $\alpha$ was fixed to be $\alpha = 10$.
}
\label{table:Tc_adj}
\end{table}

\begin{table}[ht]
\begin{center}
\begin{tabular*}{7cm}{@{\extracolsep{\fill}}cccc}
\hline
$a_0$  & $a_1$   & $a_2$   & $b_3$ \\
\hline
$3.51$ & $-2.47$ & $15.22$ & $-1.75$ \\
\hline
\end{tabular*}
\end{center}
\caption{ Set of parameters for the improved Polyakov-loop effective
potential~\cite{RRW:phase}. }
\label{table:imp}
\end{table}

\end{widetext}


\end{document}